\documentclass[%
reprint,
superscriptaddress,
nofootinbib,
amsmath,amssymb,
aps,prc
]{revtex4-2}

\usepackage{caption}
\usepackage{subcaption}
\usepackage{graphicx}
\usepackage{dcolumn}
\usepackage{bm}
\usepackage{xcolor}
\usepackage{float}
\usepackage{footnote}
\usepackage{mathtools}

\usepackage{xr-hyper}
\usepackage{hyperref}

\usepackage{url}
\usepackage{multirow}
\usepackage{xfrac}

\begin{document}

\title{Simultaneous $\gamma$-ray and electron spectroscopy of $^{182,184,186}$Hg isotopes}
\author{M.~Stryjczyk}
\email{marek.m.stryjczyk@jyu.fi, he/him}
\affiliation{KU Leuven, Instituut voor Kern- en Stralingsfysica, Celestijnenlaan 200D, 3001 Leuven, Belgium}
\affiliation{University of Jyvaskyla, Department of Physics, Accelerator laboratory, P.O. Box 35, FI-40014 University of Jyvaskyla, Finland}
\author{B.~Andel}
\thanks{These authors contributed equally}
\affiliation{KU Leuven, Instituut voor Kern- en Stralingsfysica, Celestijnenlaan 200D, 3001 Leuven, Belgium}
\affiliation{Department of Nuclear Physics and Biophysics, Comenius University in Bratislava, 84248 Bratislava, Slovakia}
\author{J.~G.~Cubiss} 
\thanks{These authors contributed equally}
\affiliation{ISOLDE, CERN, CH-1211 Geneva 23, Switzerland}
\affiliation{School of Physics, Engineering and Technology, University of York, York, YO10 5DD, United Kingdom}
\author{K.~Rezynkina}
\thanks{These authors contributed equally}
\affiliation{KU Leuven, Instituut voor Kern- en Stralingsfysica, Celestijnenlaan 200D, 3001 Leuven, Belgium}
\affiliation{INFN Sezione di Padova, I-35131 Padova, Italy}
\author{T.~R.~Rodr\'{\i}guez}
\affiliation{Departamento de F\'isica Te\'orica and Centro de Investigaci\'on Avanzada en F\'isica Fundamental, Universidad Aut\'onoma de Madrid, E-28049 Madrid, Spain}
\affiliation{Grupo de F\'isica Nuclear, Universidad Complutense de Madrid, 28040, Madrid, Spain}
\author{J.~E.~Garc\'ia-Ramos} 
\affiliation{Departamento de Ciencias Integradas y Centro de Estudios Avanzados en F\'isica, Matem\'atica y Computaci\'on, Universidad de Huelva, 21071 Huelva, Spain}
\affiliation{Instituto Carlos I de F\'isica Te\'orica y Computacional, Universidad de Granada, Fuentenueva s/n, 18071 Granada, Spain}
\author{A.~N.~Andreyev} 
\affiliation{School of Physics, Engineering and Technology, University of York, York, YO10 5DD, United Kingdom}
\affiliation{Advanced Science Research Center, Japan Atomic Energy Agency, Tokai-mura, Japan}
\author{J.~Pakarinen}
\affiliation{University of Jyvaskyla, Department of Physics, Accelerator laboratory, P.O. Box 35, FI-40014 University of Jyvaskyla, Finland}
\affiliation{Helsinki Institute of Physics, University of Helsinki, P.O. Box 64, FI-00014, Helsinki, Finland}
\author{P.~Van~Duppen}
\affiliation{KU Leuven, Instituut voor Kern- en Stralingsfysica, Celestijnenlaan 200D, 3001 Leuven, Belgium}
\author{S.~Antalic}
\affiliation{Department of Nuclear Physics and Biophysics, Comenius University in Bratislava, 84248 Bratislava, Slovakia}
\author{T.~Berry}
\affiliation{Department of Physics, University of Surrey, Guildford GU2 7XH, United Kingdom}
\author{M.~J.~G.~Borge} 
\affiliation{Instituto de Estructura de la Materia, CSIC, Serrano 113 bis, E-28006 Madrid, Spain}
\affiliation{ISOLDE, CERN, CH-1211 Geneva 23, Switzerland}
\author{C.~Clisu}
\affiliation{“Horia Hulubei” National Institute for Physics and Nuclear Engineering, RO-077125 Bucharest, Romania}
\author{D.~M.~Cox} 
\affiliation{Department of Physics, Lund University, Lund S-22100, Sweden}
\author{H.~De~Witte} 
\affiliation{KU Leuven, Instituut voor Kern- en Stralingsfysica, Celestijnenlaan 200D, 3001 Leuven, Belgium}
\author{L.~M.~Fraile}
\affiliation{Grupo de F\'isica Nuclear, Universidad Complutense de Madrid, 28040, Madrid, Spain}
\author{H.~O.~U.~Fynbo} 
\affiliation{Department of Physics and Astronomy, Aarhus University, DK-8000 Aarhus C, Denmark}
\author{L.~P.~Gaffney}
\affiliation{ISOLDE, CERN, CH-1211 Geneva 23, Switzerland}
\affiliation{Department of Physics, Oliver Lodge Laboratory, University of Liverpool, Liverpool L69 7ZE, United Kingdom}
\author{L.~J.~Harkness-Brennan} 
\affiliation{Department of Physics, Oliver Lodge Laboratory, University of Liverpool, Liverpool L69 7ZE, United Kingdom}
\author{M.~Huyse}
\affiliation{KU Leuven, Instituut voor Kern- en Stralingsfysica, Celestijnenlaan 200D, 3001 Leuven, Belgium}
\author{A.~Illana}
\affiliation{Istituto Nazionale di Fisica Nucleare, Laboratori Nazionali di Legnaro, Legnaro 35020, Italy}
\affiliation{University of Jyvaskyla, Department of Physics, Accelerator laboratory, P.O. Box 35, FI-40014 University of Jyvaskyla, Finland}
\affiliation{Grupo de F\'isica Nuclear, Universidad Complutense de Madrid, 28040, Madrid, Spain}
\author{D.~S.~Judson} 
\affiliation{Department of Physics, Oliver Lodge Laboratory, University of Liverpool, Liverpool L69 7ZE, United Kingdom}
\author{J.~Konki}
\affiliation{ISOLDE, CERN, CH-1211 Geneva 23, Switzerland}
\author{J.~Kurcewicz} 
\affiliation{ISOLDE, CERN, CH-1211 Geneva 23, Switzerland}
\author{I.~Lazarus} 
\affiliation{STFC Daresbury Laboratory, Daresbury, Warrington WA4 4AD, United Kingdom}
\author{R.~Lica}
\affiliation{“Horia Hulubei” National Institute for Physics and Nuclear Engineering, RO-077125 Bucharest, Romania}
\affiliation{ISOLDE, CERN, CH-1211 Geneva 23, Switzerland}
\author{M.~Madurga} 
\affiliation{ISOLDE, CERN, CH-1211 Geneva 23, Switzerland}
\author{N.~Marginean} 
\affiliation{“Horia Hulubei” National Institute for Physics and Nuclear Engineering, RO-077125 Bucharest, Romania}
\author{R.~Marginean} 
\affiliation{“Horia Hulubei” National Institute for Physics and Nuclear Engineering, RO-077125 Bucharest, Romania}
\author{C.~Mihai} 
\affiliation{“Horia Hulubei” National Institute for Physics and Nuclear Engineering, RO-077125 Bucharest, Romania}
\author{P.~Mosat}
\affiliation{Department of Nuclear Physics and Biophysics, Comenius University in Bratislava, 84248 Bratislava, Slovakia}
\author{E.~Nacher} 
\affiliation{Instituto de F\'isica Corpuscular, CSIC - Universidad de Valencia, E-46980, Valencia, Spain}
\author{A.~Negret} 
\affiliation{“Horia Hulubei” National Institute for Physics and Nuclear Engineering, RO-077125 Bucharest, Romania}
\author{J.~Ojala}
\affiliation{University of Jyvaskyla, Department of Physics, Accelerator laboratory, P.O. Box 35, FI-40014 University of Jyvaskyla, Finland}
\affiliation{Helsinki Institute of Physics, University of Helsinki, P.O. Box 64, FI-00014, Helsinki, Finland}
\author{J.~D.~Ovejas}
\affiliation{Instituto de Estructura de la Materia, CSIC, Serrano 113 bis, E-28006 Madrid, Spain}
\author{R.~D.~Page} 
\affiliation{Department of Physics, Oliver Lodge Laboratory, University of Liverpool, Liverpool L69 7ZE, United Kingdom}
\author{P.~Papadakis}
\affiliation{Department of Physics, Oliver Lodge Laboratory, University of Liverpool, Liverpool L69 7ZE, United Kingdom}
\affiliation{STFC Daresbury Laboratory, Daresbury, Warrington WA4 4AD, United Kingdom}
\author{S.~Pascu} 
\affiliation{“Horia Hulubei” National Institute for Physics and Nuclear Engineering, RO-077125 Bucharest, Romania}
\author{A.~Perea} 
\affiliation{Instituto de Estructura de la Materia, CSIC, Serrano 113 bis, E-28006 Madrid, Spain}
\author{Zs.~Podoly\'ak}
\affiliation{Department of Physics, University of Surrey, Guildford GU2 7XH, United Kingdom}
\author{L.~Pr\'ochniak} 
\affiliation{Heavy Ion Laboratory, University of Warsaw, PL-02-093, Warsaw, Poland}
\author{V.~Pucknell} 
\affiliation{STFC Daresbury Laboratory, Daresbury, Warrington WA4 4AD, United Kingdom}
\author{E.~Rapisarda} 
\affiliation{ISOLDE, CERN, CH-1211 Geneva 23, Switzerland}
\author{F.~Rotaru} 
\affiliation{“Horia Hulubei” National Institute for Physics and Nuclear Engineering, RO-077125 Bucharest, Romania}
\author{C.~Sotty}
\affiliation{“Horia Hulubei” National Institute for Physics and Nuclear Engineering, RO-077125 Bucharest, Romania}
\author{O.~Tengblad} 
\affiliation{Instituto de Estructura de la Materia, CSIC, Serrano 113 bis, E-28006 Madrid, Spain}
\author{V.~Vedia} 
\affiliation{Grupo de F\'isica Nuclear, Universidad Complutense de Madrid, 28040, Madrid, Spain}
\author{S.~Vi\~nals} 
\affiliation{Instituto de Estructura de la Materia, CSIC, Serrano 113 bis, E-28006 Madrid, Spain}
\author{R.~Wadsworth} 
\affiliation{School of Physics, Engineering and Technology, University of York, York, YO10 5DD, United Kingdom}
\author{N.~Warr} 
\affiliation{Institut f\"{u}r Kernphysik, Universit\"{a}t zu K\"{o}ln, 50937 K\"{o}ln, Germany}
\author{K.~Wrzosek-Lipska}
\affiliation{Heavy Ion Laboratory, University of Warsaw, PL-02-093, Warsaw, Poland}

\collaboration{IDS Collaboration}

\date{\today}

\begin{abstract}
\begin{description}
\item[Background] The mercury isotopes around $N=104$ are a well-known example of nuclei exhibiting shape coexistence. Mixing of configurations can be studied by measuring the monopole strength $\rho^2(E0)$, however, currently the experimental information is scarce and lacks precision, especially for the $I^\pi \rightarrow I^\pi$ ($I \neq 0$) transitions.
\item[Purpose] The goals of this study were to increase the precision of the known branching ratios and internal conversion coefficients, to increase the amount of available information regarding excited states in $^{182,184,186}$Hg and to interpret the results in the framework of shape coexistence using different models.
\item[Method] The low-energy structures in $^{182,184,186}$Hg were populated in the $\beta$ decay of $^{182,184,186}$Tl, produced at ISOLDE, CERN and purified by laser ionization and mass separation. The $\gamma$-ray and internal conversion electron events were detected by five germanium clover detectors and a segmented silicon detector, respectively, and correlated in time to build decay schemes.  
\item[Results] In total, 193, 178 and 156 transitions, including 144, 140 and 108 observed for the first time in a $\beta$-decay experiment, were assigned to $^{182,184,186}$Hg, respectively. Internal conversion coefficients were determined for 23 transitions, out of which 12 had an $E0$ component. Extracted branching ratios allowed the sign of the interference term in $^{182}$Hg as well as $\rho^2(E0;0^+_2\rightarrow 0^+_1)$ and $B(E2;0^+_2\rightarrow 2^+_1)$ in $^{184}$Hg to be determined. By means of electron-electron coincidences, the $0^+_3$ state was identified in $^{184}$Hg. The experimental results were qualitatively reproduced by five theoretical approaches, the Interacting Boson Model with configuration mixing with two different parametrizations, the General Bohr Hamiltonian, the Beyond Mean-Field model and the symmetry-conserving configuration-mixing model. However, a quantitative description is lacking. 
\item[Conclusions] The presence of shape coexistence in neutron-deficient mercury isotopes was confirmed and evidence for the phenomenon existing at higher energies was found. The new experimental results provide important spectroscopic input for future Coulomb excitation studies.
\end{description}
\end{abstract}

\maketitle

\section{\label{sec:introduction}Introduction}

The neutron-deficient mercury isotopes ($Z=80$) around the neutron mid-shell at $N=104$ constitute one of the most prominent examples of shape coexistence \cite{Heyde2011}. Laser spectroscopy studies in this region show dramatic changes of the charge radii between the neighboring isotopes \cite{Marsh2018,Sels2019}. This behavior, called shape staggering, indicates a large change of deformation between the measured ground and isomeric states \cite{Heyde2011}. The evolution of shape coexistence is demonstrated in the level energy systematics of the even-mass mercury isotopes that show two structures at low energies, one built on top of the ground state, interpreted as weakly oblate-deformed, and the other, built on top of the intruder $0^+_2$ state, assumed to be prolate-deformed \cite{Julin2001,Heyde2011}. The excitation energies of the latter have a parabolic behavior as a function of neutron number, with the minimum at $N=102$ in $^{182}$Hg.

The presence of two coexisting bands is confirmed by other complementary experiments in this region. Lifetime measurements of the yrast-band members up to the $8^+$ state in even-mass $^{180-188}$Hg isotopes have shown large $E2$ transition strengths, while they drop for the $2^+_1$ state \cite{Grahn2009,Scheck2010,Gaffney2014,Siciliano2020}. This behavior indicates a similar configuration of high-spin states and a mixing of two configurations in the $2^+_1$ level. One should also note the decrease of the $4^+_1 \rightarrow 2^+_1$ transition strength from $^{180}$Hg, where it is similar to the values between higher-spin members of the yrast cascade \cite{Grahn2009}, to $^{188}$Hg, where it is much closer to the $B(E2;2^+_1 \rightarrow 0^+_1)$ \cite{Siciliano2020}. This effect is interpreted as an evolution of the $4^+_1$ state structure from prolate- to oblate-deformed shape \cite{Grahn2009}.

The $^{186,188,190}$Pb $\alpha$-decay fine structure measurements reveal large hindrance factors for decays to the $0^+_2$ states in $^{182,184,186}$Hg, which is interpreted as an indication of a weak mixing between the $0^+$ states \cite{Wauters1993,Wauters1994,Richards1996}. On the other hand, internal conversion coefficient (ICC) measurements between the first and the second $2^+$ states point to the existence of a large $E0$ component \cite{Elseviers2011,Rapisarda2017,Cole1977,Beraud1977,Scheck2011}, which is interpreted as a fingerprint of mixing \cite{Wood1999,Heyde2011}. A Coulomb excitation (Coulex) study at ISOLDE \cite{Bree2014,Wrzosek-Lipska2019} provided the monopole strengths between the lowest $2^+$ states and it confirmed strong mixing between these states. 

While the existing experimental information points to a good qualitative description of shape coexistence in mercury isotopes, quantitative information is still lacking. Currently, the insufficient precision of the spectroscopic information, with uncertainties of the $\gamma$-branching ratios and the ICCs being as large as 30\% \cite{Rapisarda2017}, hinders the interpretation of the Coulex results \cite{Wrzosek-Lipska2016}. Information on mixing of the $4^+$ and higher-spin states is also lacking. Different theoretical approaches have been tested in the region and while they are able to reproduce some of the observables, they point to contradicting conclusions, for instance regarding the intrinsic deformation of the $^{186}$Hg ground state \cite{Sels2019,Algora2021}. 

In order to increase the available amount of spectroscopic information and its precision, excited states in $^{182,184,186}$Hg have been studied by means of the $\beta$ decay of $^{182,184,186}$Tl at the ISOLDE facility at CERN. The existence of isomers in the thallium isotopes with spin and parity $2^-$, $4^-$, $7^+$ and $10^-$ \cite{Barzakh2013,Barzakh2017} enabled the population of excited states in the $^{ 182,184,186}$Hg isotopes up to spin 12 while the simultaneous detection of $\gamma$ rays and electrons allowed us to measure ICCs and, consequently, to identify transitions with $E0$ components. 

The paper is organized as follows. In Sec. \ref{sec:setup} the experimental setup is described. The analysis methods and information relevant for all three cases are presented at the beginning of Sec. \ref{sec:results} and the results for $^{182}$Hg, $^{184}$Hg and $^{186}$Hg are provided in Secs. \ref{sec:182Hg}, \ref{sec:184Hg} and \ref{sec:186Hg}, respectively. In Sec. \ref{sec:delta} a method to extract mixing ratios for all three isotopes is presented together with the results. The discussion and the interpretation of the results as well as the comparison with the theoretical calculations are given in Sec. \ref{sec:discussion}. In Sec. \ref{sec:conclusions}, conclusions are drawn and an outlook is provided. 

\section{\label{sec:setup}Experimental setup}

Pure beams of $^{182,184,186}$Tl were produced at the ISOLDE facility at CERN \cite{Catherall2017} in spallation of a thick UC$_x$ target by 1.4 GeV protons, delivered every 1.2 seconds or a multiple of this value by the Proton Synchrotron Booster (PSB). The produced nuclei diffused from the target material to a hot cavity, where the thallium isotopes were selectively ionized by the Resonance Ionization Laser Ion Source system \cite{Fedosseev2017} in a two-step ionization process. The first step excitation was performed via the 6p $^2$P$_{1/2} \rightarrow$ 6d $^2$D$_{3/2}$ transition at 276.83 nm using a dye laser system and for the second step, the Nd:YAG laser at 532 nm was used. The ionized thallium isotopes were extracted from the ion source at 30 keV energy and mass-separated by the High Resolution Separator \cite{Catherall2017}. The beam was implanted into a movable tape at the center of the ISOLDE Decay Station (IDS) \cite{IDS}. The tape was moved every 30 to 50 seconds, depending on the structure of the PSB supercycle, in order to remove daughter activities.

To detect the internal conversion electrons (ICE), the SPEDE spectrometer \cite{Papadakis2018} was employed. In its heart there is a 24-fold segmented, 1-mm thick annular silicon detector cooled by circulating ethanol at about $-20^{\mathrm{o}}\mathrm{C}$. The SPEDE spectrometer was placed inside the IDS decay chamber at 16-mm distance in front of the tape, in the upstream direction of the beam. For the detection of $\beta$ particles, a 0.5-mm thick 900 mm$^2$ silicon detector was mounted in the downstream direction. The $\gamma$ radiation was detected by five High-Purity Germanium Clover detectors (HPGe). Four of them were placed in the upstream direction while the fifth one was placed in the downstream direction and it was used only for energy gating. Signals from the detectors were recorded using the Nutaq digital data acquisition system \cite{Nutaq} with 100 MHz sampling frequency, running in a triggerless mode. 

To calibrate the germanium detectors, an encapsulated $^{152}$Eu source and a $^{138}$Cs sample, produced on-line and implanted onto the tape, were used, while for the SPEDE spectrometer, the ICEs from the strong $E2$ transitions in $^{184,186}$Hg and $^{138}$Ba were utilized. More details regarding the setup calibration and its performance are reported in Refs. \cite{thesis,StryjczykSPEDE}.  

\section{\label{sec:results}Results}

The $\beta$-decay schemes of $^{182,184,186}$Tl were built using $\gamma$-$\gamma$, $\gamma$-electron and electron-electron coincidence spectra. The coincidence time window between any two signals was 300~ns. In all three measured cases, the beam was a mixture containing two or three $\beta$-decaying isomers of thallium in an unknown proportion. As a result, only $\gamma$-branching ratios for each excited state were extracted while apparent $\beta$ feedings and log(\textit{ft}) values were not determined. Tables with $\gamma$-branching ratios, $\gamma$-ray intensities normalized to the strongest $2^+_1 \rightarrow 0^+_1$ transition for each isotope and full decay schemes are provided in the Supplemental Material \cite{supplemental}. In the following sections, the information relevant for each isotope is presented.

\subsection{\label{sec:182Hg}Excited states in $^{182}$Hg}

\begin{figure}
\includegraphics[width=\columnwidth]{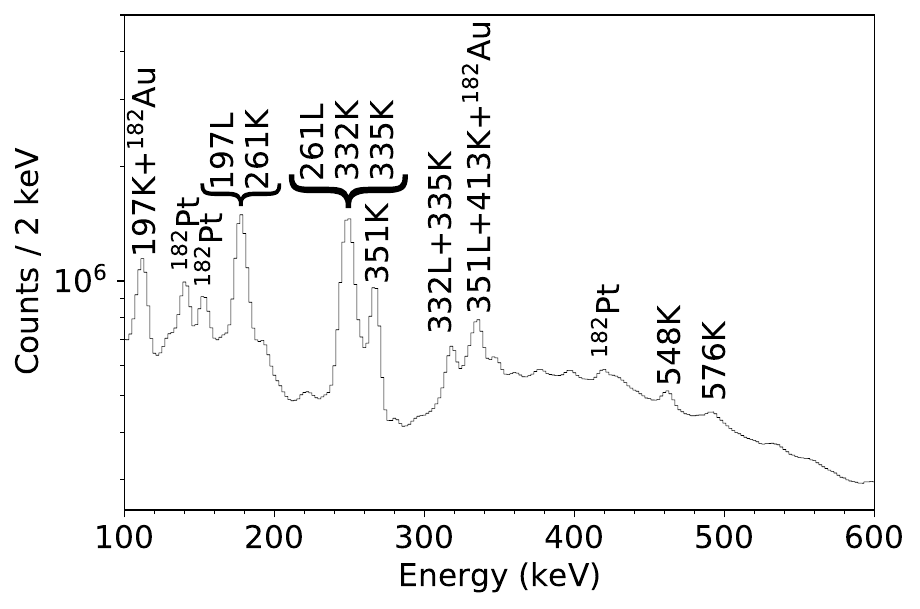}
\caption{\label{fig:182Hg_e}Portion of the electron singles energy spectrum collected with the $^{182}$Tl beam. The electron lines associated with $^{182}$Hg have been marked with the transition energy and corresponding atomic orbital, whereas transitions arising from the $A=182$ decay chain are marked with the nucleus of origin.}
\end{figure}

\begin{figure}
\includegraphics[width=\columnwidth]{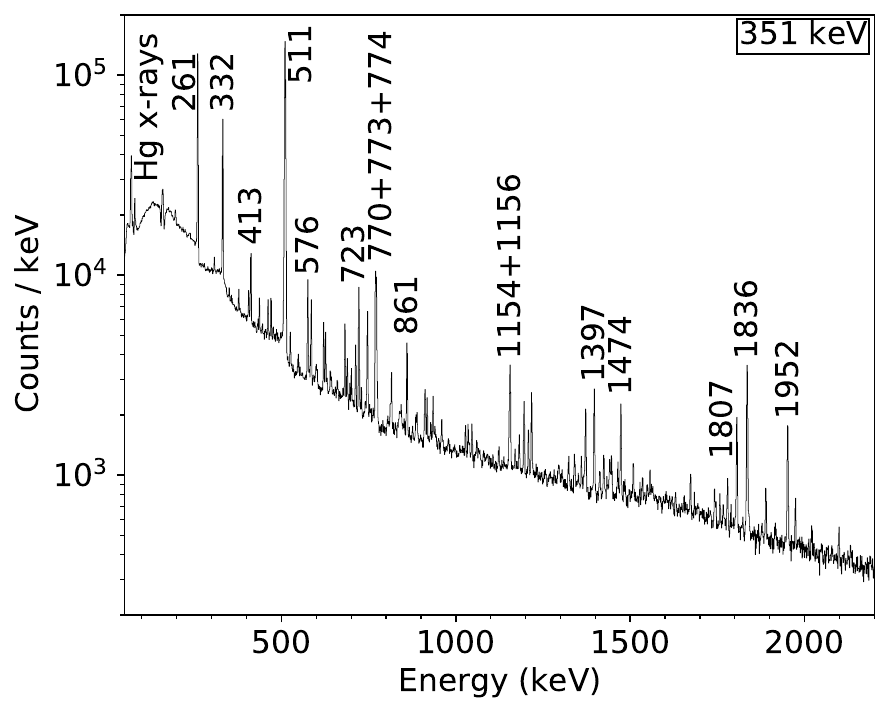}
\caption{\label{fig:182Hg_gg}Portion of the $\gamma$-ray energy spectrum gated on the 351-keV ($2^+_1 \rightarrow 0^+_1$) $\gamma$-ray transition in $^{182}$Hg. The most prominent lines have been labeled with the transition energy in keV.}
\end{figure}

\begin{figure*}
\includegraphics[width=0.8\textwidth]{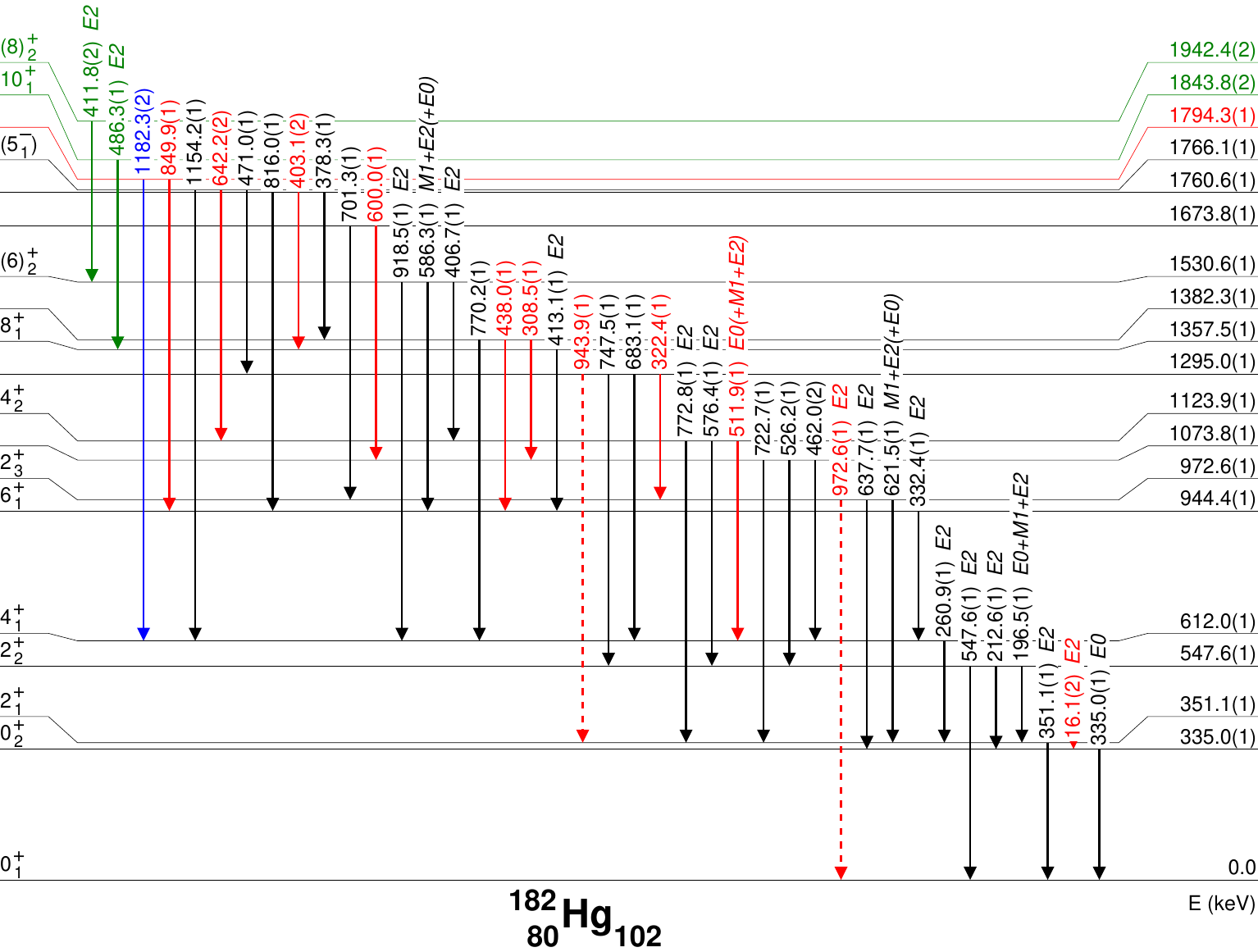}
\caption{\label{fig:182Hgdecayscheme}Partial level scheme of excited states in $^{182}$Hg populated in the $\beta$ decay of $^{182}$Tl extracted in this work. For the full version, see Supplemental Material \cite{supplemental}. Levels and transitions known from the previous $\beta$-decay studies are plotted in black, shifted in the decay scheme in blue, known from other than $\beta$-decay studies in green and newly identified in red. Transitions not observed in this work for which the intensity limits have been determined are plotted with dashed lines. Spins, parities and proposed transition multipolarities are taken from this work and Ref. \cite{Singh2015}.}
\end{figure*}

The analysis of the coincidence data allowed us to confirm the decay scheme proposed by Rapisarda \textit{et al.} in Ref. \cite{Rapisarda2017}. The only exception was the 1182-keV transition which was moved from the 2566-keV state to the 1794-keV state. Electron singles energy spectrum and a typical $\gamma$-ray energy spectrum with a gate on a $\gamma$ ray are presented in Figs. \ref{fig:182Hg_e} and \ref{fig:182Hg_gg}. In total, 89 excited states and 193 transitions were identified  in $^{182}$Hg. Out of them, there were 57 new excited states and 136 new transitions. Six levels and eight transitions known from in-beam studies \cite{Bindra1995} were also observed. It should be noted that we observed a systematic shift of around 1 keV between the $\gamma$-ray energies reported in our work and Ref. \cite{Bindra1995}. A similar shift was observed in the previous $\beta$-decay study \cite{Rapisarda2017}. In addition, nine ICCs have been measured. A summary of the deduced levels with their de-exciting transitions is presented in Tables I and II in Supplemental Material \cite{supplemental}. The ICCs are given in Table \ref{tab:conversion182Hg} and a partial decay scheme is shown in Fig. \ref{fig:182Hgdecayscheme}.

\begin{table*}[h!t!b]
\caption{\label{tab:conversion182Hg}
Experimental internal conversion coefficients $\alpha_{exp}$ of transitions in $^{182}$Hg compared with the theoretical values $\alpha_{th}$ calculated using BrIcc \cite{Kibedi2008} and the deduced transition multipolarities $X\lambda$.}
\begin{ruledtabular}
\begin{tabular}{ccccccccc}
$E_{i}$ (keV) 		& $E_{f}$ (keV)	& $E_t$ (keV)			& Transition 		& Shell 		& $\alpha_{exp}$ 			& $\alpha_{th}(E2)$ 	& $\alpha_{th}(M1)$ & $X\lambda$\\\hline
547.6(1) 	& 351.1(1) 	& 196.5(1) 	& $2^+_2 \rightarrow 2^+_1$ 	& K & 6.0(7) 	& 0.1768(25) 	& 0.952(14) & E0+M1+E2\\
			&  			& 			& 								& L & 1.24(15) 	& 0.179(3) 	 	& 0.1602(23) & \\
			&  			&  			& 								& M+ &0.37(5) 	& 0.0596(7) 	& 0.0486(6) & \\
547.6(1) 	& 335.0(1)	& 212.6(1) 	& $2^+_2 \rightarrow 0^+_2$ 	& L & 0.132(27) 	& 0.1285(19) 	& 0.1285(18) & E2\\
547.6(1) 	& 0.0 		& 547.6(1) 	& $2^+_2 \rightarrow 0^+_1$ 	& K & 0.015(2) 	& 0.01603(23) 	& 0.0595(9) & E2\\
972.6(1)  	& 351.1(1) 	& 621.5(1) 	& $2^+_3 \rightarrow 2^+_1$ 	& K & 0.045(13)	& 0.01231(18) 	& 0.0428(6) & M1+E2(+E0)\\
972.6(1)  	& 335.0(1)	& 637.7(1) 	& $2^+_3 \rightarrow 0^+_2$ 	& K & $<$ 0.029\footnotemark[1] 	& 0.01168(17) 	& 0.0400(6) & E2\\
1123.9(1) 	& 612.0(1)	& 511.9(1)	& $4^+_2 \rightarrow 4^+_1$ 	& all & $>$ 0.65\footnotemark[1] 	& 0.0255(4) 	& 0.0862(12) & E0(+M1+E2)\\
1530.6(1) 	& 944.4(1) 	& 586.3(1)	& $(6)^+_2 \rightarrow 6^+_1$ 	& K & 0.030(8) 	& 0.01389(2) 	& 0.0498(7) & M1+E2(+E0) \\
1718.5(1)	& 1507.4(1)	& 211.0(2)	& 							    & K & $>$ 0.90\footnotemark[1] 	& 0.1489(22) & 0.781(12) & E0(+M1+E2)\\
1984.9(1)	& 1766.1(1)	& 219.1(2)	& $(5^-_2) \rightarrow (5^-_1)$	& K & 0.90(21)	& 0.1358(20)	& 0.703(10)	& M1\\
\end{tabular}
\end{ruledtabular}
\footnotetext[1]{Limit given with 95\% credible interval.}
\end{table*}
\begin{figure}
\includegraphics[width=\columnwidth]{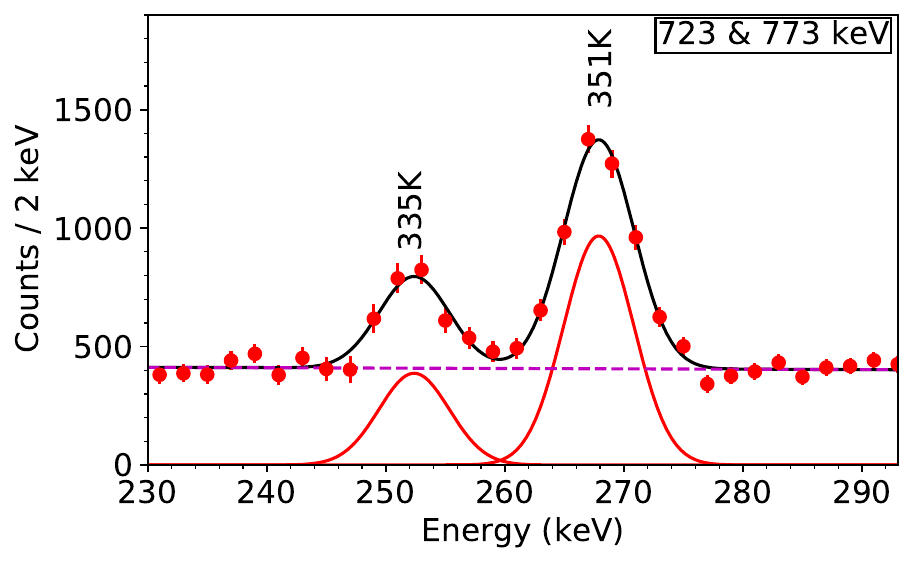}
\caption{\label{fig:gefit_335_351}A fit (black solid line) to the portion of the electron energy spectrum spectrum, gated on the 723- and 773-keV $\gamma$ rays in $^{182}$Hg. The visible peaks stem from the K-ICE of the 335-keV ($0^+_2 \rightarrow 0^+_1$) and 351-keV ($2^+_1 \rightarrow 0^+_1$) transitions. The contribution from asymmetric Gaussian functions is plotted in solid red lines and the constant background in dashed purple line.}
\end{figure}

The electron energy spectrum gated on the 723 and 773 keV $\gamma$ rays feeding the $2^+_1$ 351-keV state is presented in Fig. \ref{fig:gefit_335_351}. Two peaks are visible at 268 and 252 keV, which can be associated with the K-ICE from the 351-keV $2^+_1 \rightarrow 0^+_1$ transition and the de-excitation of the 335-keV $0^+_2$ state, respectively. This observation proves the existence of a 16-keV $2^+_1 \rightarrow 0^+_2$ transition. The intensity ratio of these two peaks can be linked to the $\gamma$-ray intensity ratio de-exciting the 351-keV state:
\begin{equation}
\frac{I_\gamma(16)}{I_\gamma(351)} = \frac{I_K(335)}{I_K(351)} \frac{\Omega_{tot}(335)}{\Omega_K(335)} \frac{\alpha_K(351)}{1+\alpha_{tot}(16)} = 5.8(7) \times 10^{-7} \mathrm{,}
\end{equation}
where $\Omega_K(335)$ and $\Omega_{tot}(335)$ are the tabulated K and total electronic factors for the 335-keV $E0$ transitions, respectively, taken from Ref. \cite{Dowie2020}, while $\alpha_K(351)$ and $\alpha_{tot}(16)$ are the calculated K-ICC of the 351-keV $E2$ transition and the total ICC of the 16-keV $E2$ transitions, respectively \cite{Kibedi2008}.

The extracted value can be converted into the ratio of the $B(E2)$ transition strengths:
\begin{equation}
\frac{B(E2;16)}{B(E2;351)} = \frac{I_\gamma(16)}{I_\gamma(351)} \times \frac{E_\gamma^5(351)}{E_\gamma^5(16)} = 2.9(4) \mathrm{.}
\end{equation}
Having this ratio and the $B(E2;2^+_1 \rightarrow 0^+_1) = 0.33(2) e^2b^2$ value obtained in the Coulomb excitation studies \cite{Wrzosek-Lipska2019}, the absolute value of the matrix element ${|\langle 0^+_2\parallel E2\parallel 2^+_1\rangle}| = 2.2(3) eb$ was extracted. This result is in agreement with the $[-2.2, 0.9]$ range given in Ref. \cite{Wrzosek-Lipska2019} but only for the negative values. It is also in a good agreement with $\langle 0^+_2\parallel E2\parallel 2^+_1\rangle = -2.48 eb$ from the two-state mixing calculations presented in Ref. \cite{Gaffney2014} (see also Fig. 16 in Ref. \cite{Wrzosek-Lipska2019}).

Although the sign of an individual reduced matrix element has no physical meaning and depends solely on the used convention, the sign of the interference term is an experimental observable. It is a product of three reduced matrix elements and it is important in the determination of the state's triaxiality using the quadrupole sum rule \cite{Kumar1972,Cline1986,Wrzosek-Lipska2012}. The combined analysis of this work and the results from Ref. \cite{Wrzosek-Lipska2019} yields a sign of the $\langle 0^+_2\parallel E2\parallel 2^+_1\rangle\langle 2^+_1\parallel E2\parallel 2^+_2\rangle\langle 2^+_2\parallel E2\parallel 0^+_2\rangle$ interference term to be negative.

\begin{figure}
\includegraphics[width=\columnwidth]{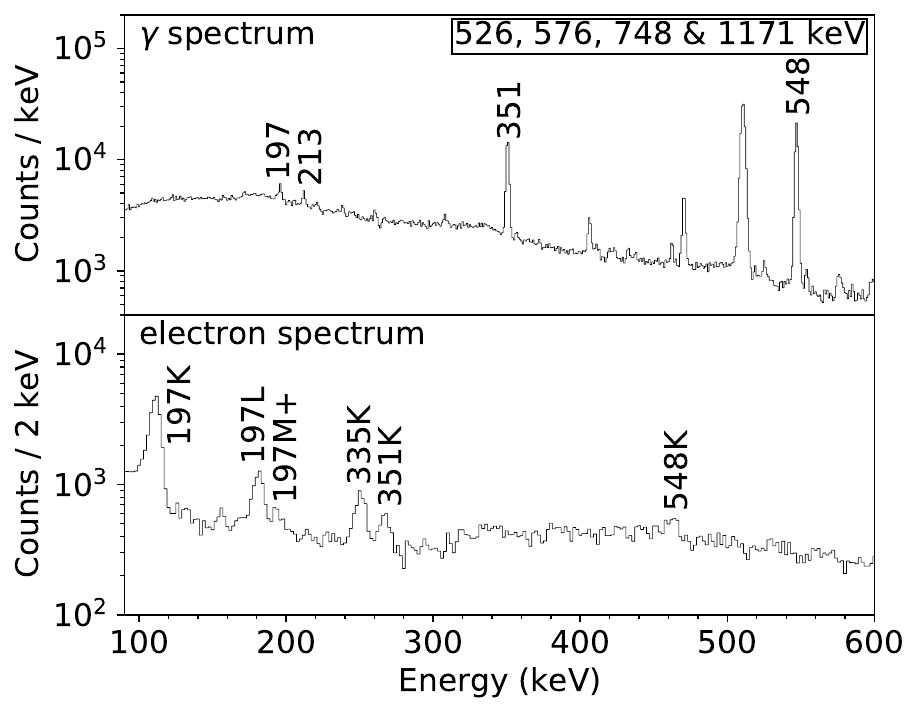}
\caption{\label{fig:feedingto548}$\gamma$-ray (top) and electron (bottom) energy spectra gated on the 526-, 576-, 748- and 1171-keV $\gamma$-ray transitions feeding the $2^+_2$ state in $^{182}$Hg. The peaks of interest are labeled by the energy given in keV.}
\end{figure}

\begin{figure}
\includegraphics[width=\columnwidth]{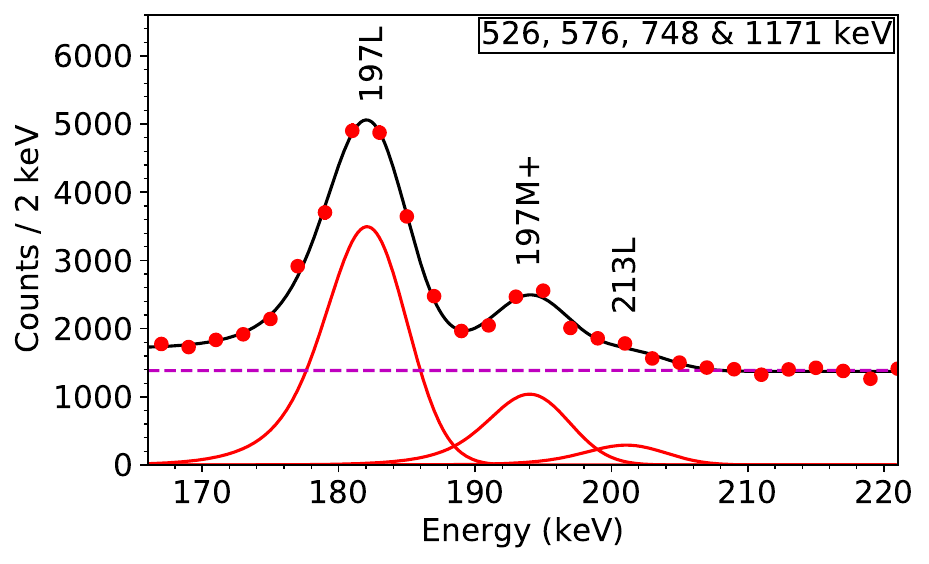}
\caption{\label{fig:182Hg_197LM_fit}A fit (black solid line) to a portion of the electron energy spectrum spectrum, gated on the 526-, 576-, 748- and 1171-keV $\gamma$ rays feeding the 548 keV $2^+_2$ state in $^{182}$Hg. The visible peaks stem from the L- and M+-ICE of the 197-keV transition and the L-ICE of the 213-keV transition. A contribution from asymmetric Gaussian functions is plotted in solid red lines and the constant background is shown by the dashed purple line.}
\end{figure}

The K-, L- and M+\footnote{Throughout this publication, M+ means electrons from M and higher atomic shells. The energy resolution of the SPEDE Spectrometer does not allow electrons stemming from these atomic shells to be resolved.}-internal conversion coefficients of the ${2^+_2 \rightarrow 2^+_1}$ transition were determined from the $\gamma$-ray and electron energy spectra gated on the 526-, 576-, 748- and 1171-keV $\gamma$ rays (see Figs. \ref{fig:feedingto548} and \ref{fig:182Hg_197LM_fit}). A fit to the L and M+ electrons is presented in Fig. \ref{fig:182Hg_197LM_fit}. The sum of the extracted ICCs, which is equal to 7.6(7), is in a good agreement with the value of 7.2(13) reported in Ref. \cite{Rapisarda2017}. 

Employing the same gate, the K-ICC of the 548-keV ${2^+_2 \rightarrow 0^+_1}$ transition and the L-ICC of the 213-keV ${2^+_2 \rightarrow 0^+_2}$ transition were extracted. Both results are in excellent agreement with the theoretical value for $E2$ transitions \cite{Kibedi2008}.

\begin{figure}
\includegraphics[width=\columnwidth]{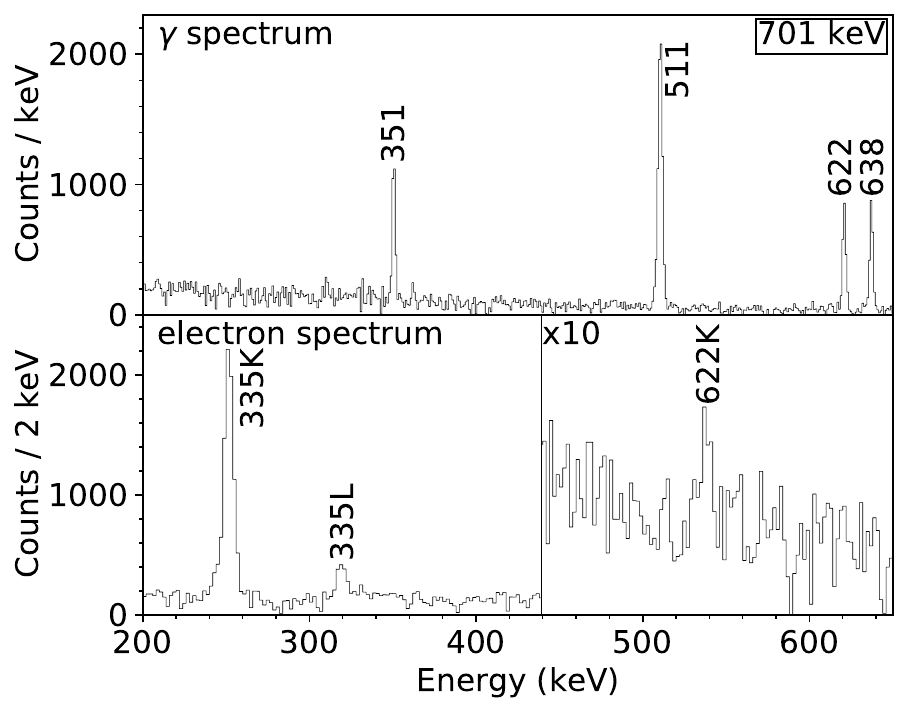}
\caption{\label{fig:622keV}$\gamma$-ray (top) and electron (bottom) energy spectra gated on the 701-keV $\gamma$ ray in $^{182}$Hg. The peaks of interest are labeled by the energy given in keV. The energy range 440-650 keV of electron energy spectrum has been magnified by a factor 10 for visualisation purposes.}
\end{figure}

The K-ICC of the 622-keV transition de-exciting the 973-keV state was obtained by gating on the 701-keV $\gamma$ ray (see Fig. \ref{fig:622keV}). Its value fixes a positive parity to the 973-keV state. The upper limit for the K-ICC of the 638-keV transition ($\alpha_K < 0.029$), extracted by employing the same gate, is consistent with a pure $E2$ character ($\alpha_K(E2) = 0.012$) and excludes an $M1$ multipolarity ($\alpha_K(M1) = 0.040$). Therefore, by combining both results, we propose the spin-parity assignment of $2^+$ for the 973-keV state. 

The energy gate set on the yrast 332-keV ($6^+_1 \rightarrow 4^+_1$) transition allowed us to extract $\alpha_K = 0.030(8)$ for the 586-keV $\gamma$ ray. This value suggests a mixed $E2/M1$ character, however, an $E0$ component cannot be excluded without an independent measurement of the $\delta$ mixing ratio. Based on this information, the de-excitation pattern and the level energy systematics (see Sec. \ref{sec:spins}), we propose spin and parity of (6)$^+$ for the 1531-keV state.

\begin{figure}
\includegraphics[width=\columnwidth]{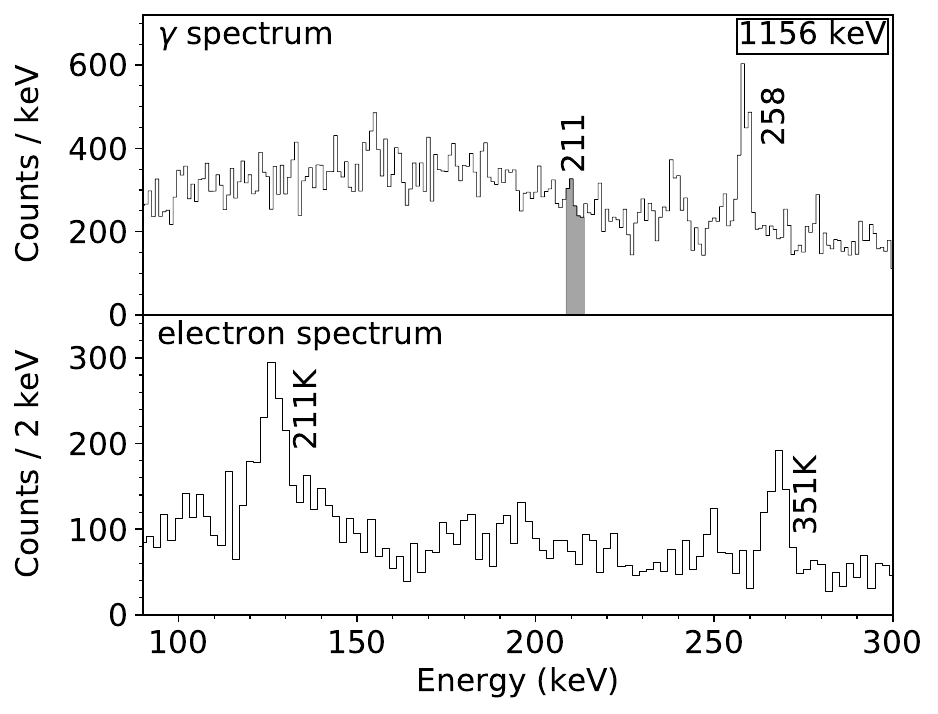}
\caption{\label{fig:211K}$\gamma$-ray (top) and electron (bottom) energy spectra gated on the 1156-keV $\gamma$ ray in $^{182}$Hg. The position of the non-observed 211-keV $\gamma$ ray is indicated with a shaded area.}
\end{figure}

The 211-keV transition de-exciting the 1719-keV state was observed only via ICEs (see Fig. \ref{fig:211K} and the decay scheme in Supplemental Materials \cite{supplemental}). The lower limit of the K-ICC ($\alpha_K > 0.9$) was extracted from the $\gamma$-ray and electron energy spectra gated on the 1156-keV $\gamma$ ray and it indicates an existence of an $E0$ component. This implies that both excited states, at 1719 and 1507 keV, have the same spin and parity.

The K-ICC of the 219-keV transition de-exciting the 1985 keV state was extracted by gating on the 576-keV $\gamma$ ray (see the decay scheme in Supplemental Materials \cite{supplemental}). The value of 0.90(21) is in 1$\sigma$ agreement with a pure $M1$ transition. By combining this information, the de-excitation of the 1985-keV state to the $4^+$ and $6^+$ states and the $(5^-)$ assignment of the 1766-keV level fed by the 219-keV transition, we propose $(5^-)$ spin-parity for the 1985-keV state.

\begin{figure}
\includegraphics[width=\columnwidth]{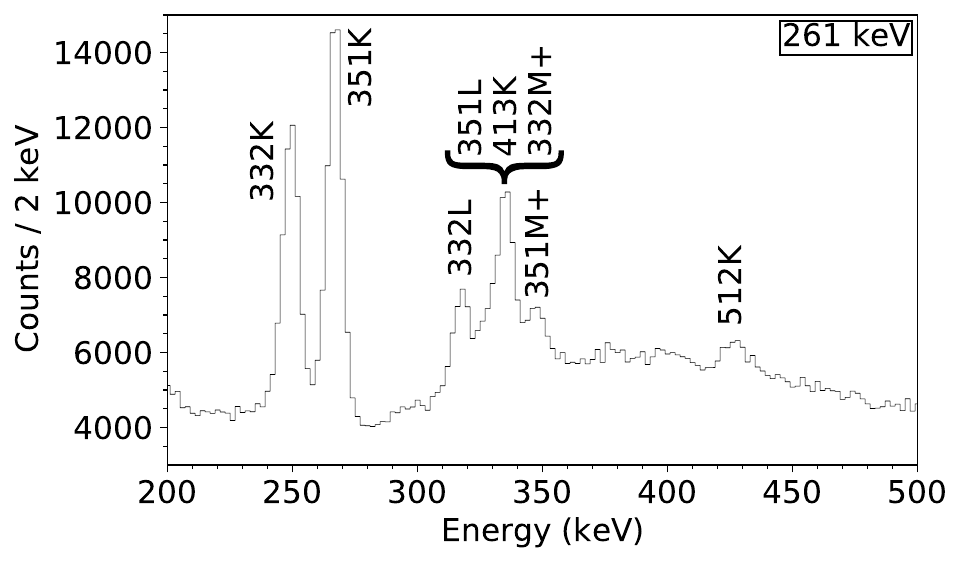}
\caption{\label{fig:511K}Portion of an electron energy spectrum gated on the 261-keV ($4^+_1 \rightarrow 2^+_1$) $\gamma$ ray in $^{182}$Hg. Peaks stemming from the ICEs of the yrast cascade as well as from the 512-keV transition are labeled.}
\end{figure}

The 512-keV transition was observed in an electron energy spectrum gated on the 261-keV $\gamma$ ray (see Fig. \ref{fig:511K}) and its placement was confirmed by matching energy as well as the presence of the 1218-keV $\gamma$ ray feeding the 1124-keV level (see the decay scheme in Supplemental Materials \cite{supplemental}) in the $\gamma$-ray energy spectrum gated on the 261-keV line. Due to theoverlapping annihilation peak, the direct measurement of $\gamma$-ray intensity of the 512-keV transition could not be made. The branching ratio of 9.9(59) was determined by comparing the number of counts of the 1218-keV $\gamma$ ray registered in coincidence with the 261- and 576-keV transitions. Due to large uncertainties, only the lower limit for the total ICC ($>0.65$) of the 512-keV transition was extracted. Nevertheless, this value indicates the existence of a large $E0$ component in the 512-keV transition which allows us to firmly confirm the $4^+$ spin of the 1124-keV level.

\subsection{\label{sec:184Hg}Excited states in $^{184}$Hg}

\begin{figure}
\includegraphics[width=\columnwidth]{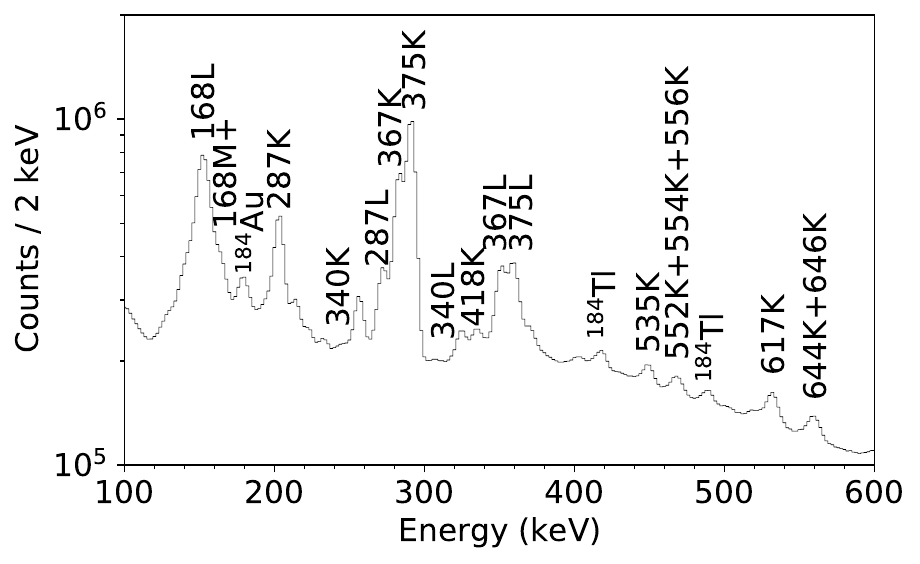}
\caption{\label{fig:184Hg_e}Portion of the electron singles energy spectrum collected with the $^{184}$Tl beam. The electron lines associated with $^{184}$Hg have been marked with the transition energy and corresponding atomic orbital, whereas transitions arising from the $A=184$ decay chain are marked with the nucleus of origin.}
\end{figure}

\begin{figure}
\includegraphics[width=\columnwidth]{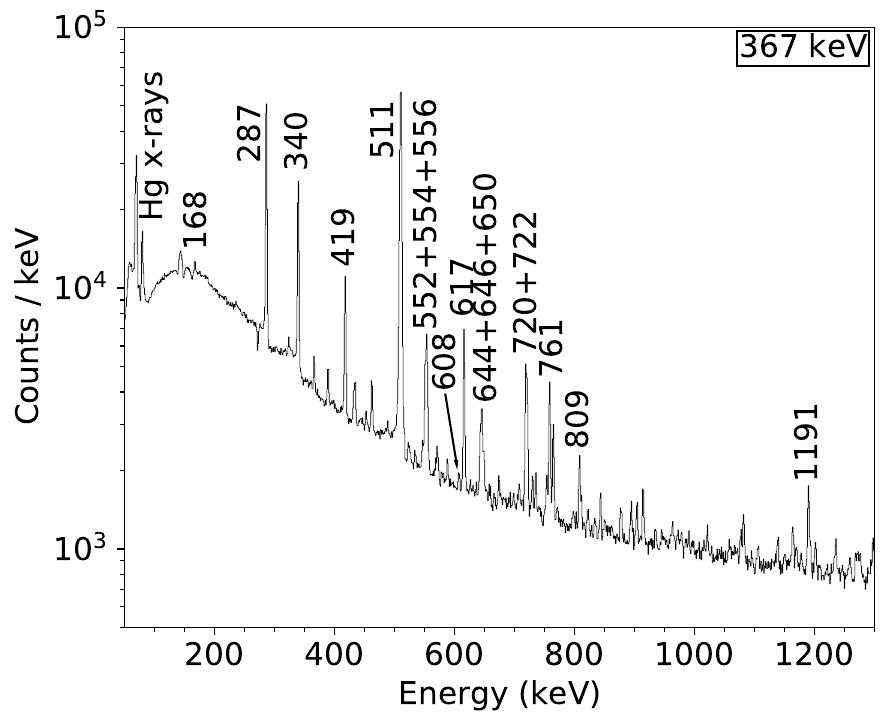}
\caption{\label{fig:184Hg_gg}Portion of the $\gamma$-ray energy spectrum gated on the 367-keV ($2^+_1 \rightarrow 0^+_1$) $\gamma$-ray transition in $^{184}$Hg. The most prominent lines have been labeled with the transition energy in keV.}
\end{figure}

\begin{figure}
\includegraphics[width=\columnwidth]{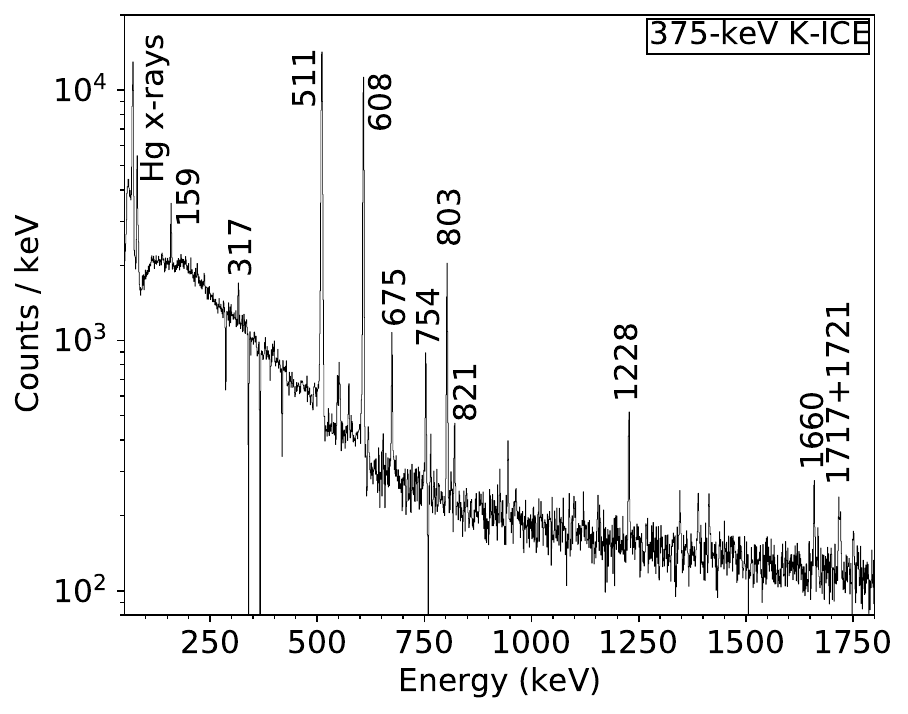}
\caption{\label{fig:184Hg_ge}Portion of the $\gamma$-ray energy spectrum gated on the 375-keV ($0^+_2 \rightarrow 0^+_1$) K-ICE in $^{184}$Hg. The negative peaks are stemming from the background subtraction. Main coincident lines are labeled by the energy given in keV.}
\end{figure}

\begin{figure*}
\includegraphics[width=0.8\textwidth]{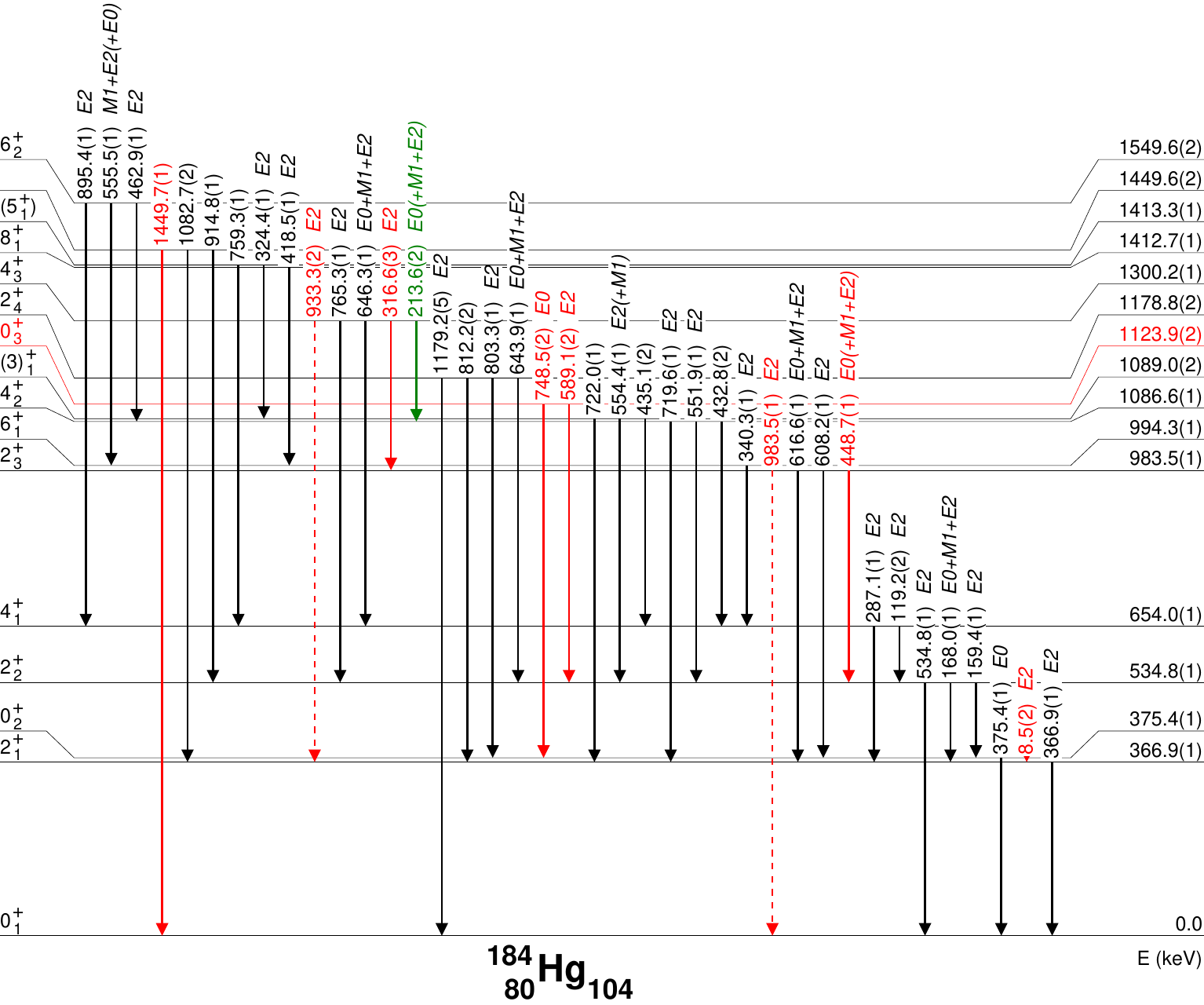}
\caption{\label{fig:184Hgdecayscheme}Partial level scheme of excited states in $^{184}$Hg populated in the $\beta$ decay of $^{184}$Tl extracted in this work. For the full version, see Supplemental Material \cite{supplemental}. Levels and transitions known from the previous $\beta$-decay studies are plotted in black, known from other studies in green and the newly identified ones in red. Transitions not observed in this work for which the intensity limits have been determined are plotted with dashed lines. Spins, parities and proposed transition multipolarities are taken from this work and Refs. \cite{Baglin2010,Rapisarda2017}.}
\end{figure*}

Based on the coincidence analysis, we confirm the decay scheme reported in Ref. \cite{Rapisarda2017}. An electron singles energy spectrum and typical $\gamma$-$\gamma$ and $\gamma$-electron spectra are presented in Figs. \ref{fig:184Hg_e}, \ref{fig:184Hg_gg} and \ref{fig:184Hg_ge}, respectively. In total, 110 excited states and 178 transitions were assigned to $^{184}$Hg. In particular, there were 126 new transitions and 85 new excited states. Four levels and 14 transitions previously observed in the in-beam studies \cite{Deng1995} were also observed in this $\beta$-decay study. Furthermore, 12 ICCs were measured. The experimental results are summarized in Table \ref{tab:conversion184Hg} and in Supplemental Material in Tables III and IV \cite{supplemental} while the partial decay scheme is presented in Fig. \ref{fig:184Hgdecayscheme}. 

\begin{table*}
\caption{\label{tab:conversion184Hg}\
ms{Experimental internal conversion coefficients $\alpha_{exp}$ of transitions in $^{184}$Hg compared with the theoretical values $\alpha_{th}$ calculated using BrIcc \cite{Kibedi2008} and the deduced transition multipolarities $X\lambda$.}}
\begin{ruledtabular}
\begin{tabular}{cccccccccc}
$E_{i}$ (keV) 	& $E_{f}$ (keV)	& $E_t$ (keV)			& Transition 					& Shell & $\alpha_{exp}$ 			& $\alpha_{th}(E2)$ 	& $\alpha_{th}(M1)$ & $X\lambda$ \\\hline
534.8(1) 		& 366.9(1)		& 168.0(1) 			& $2^+_2 \rightarrow 2^+_1$	& all & 12.8(24)\footnotemark[1] 			& 0.724(11) 			& 1.80(3) 	& E0+M1+E2\\ 
			&  				&  					& 							& K & 9.4(24)\footnotemark[2]				& 0.256(4) 			& 1.478(21) 	\\ 
			&  				&  					& 							& L & 2.6(4)			& 0.351(5) 			& 0.249(4) 	\\ 
			&  				&  					& 							& M+ & 0.78(12) 			& 0.1175(14) 			& 0.0755(9) 	\\ 
983.5(1) 		& 534.8(1) 		& 448.7(1) 			& $2^+_3 \rightarrow 2^+_2$	& K & $>$ 0.355\footnotemark[3]  & 0.0247(4) 		& 0.1005(14) 	& E0(+M1+E2) \\ 
983.5(1) 		& 366.9(1)		& 616.6(1) 			& $2^+_3 \rightarrow 2^+_1$	& K & 0.066(6) 	 			& 0.01252(18) 		& 0.0436(7) 	& E0+M1+E2\\ 
1123.9(2)	& 375.4(1) 		& 748.5(2) 			& $0^+_3 \rightarrow 0^+_2$ 	& K & $>$ 1.256\footnotemark[3]  & 0.00847(12) 		& 0.0264(4) 	& E0\\
1089.0(2) 	& 534.8(1) 		& 554.4(1) 			& $(3)^+_1 \rightarrow 2^+_2$	& K & 0.014(4)\footnotemark[4] 	 			& 0.01563(22) 		& 0.0576(8) & E2(+M1)	\\
1178.8(2) 	& 534.8(1) 		& 643.9(1)			& $2^+_4 \rightarrow 2^+_2$ 	& K & 0.100(14) 			& 0.01145(16) 		& 0.0390(6) 	& E0+M1+E2 \\
1300.2(1)	& 1086.6(1)		& 213.6(2) 			& $4^+_3 \rightarrow 4^+_2$ 	& K & $>$ 0.868\footnotemark[3] & 0.1445(21) 		& 0.755(11) 	& E0(+M1+E2)\\ 
1300.2(1) 	& 654.0(1)  		& 646.3(1) 			& $4^+_3 \rightarrow 4^+_1$ 	& K & 0.072(13) 	 			& 0.01137(16) 		& 0.0386(6) 	& E0+M1+E2 \\
1549.6(2) 	& 994.2(1) 		& 555.5(1) 			& $6^+_2 \rightarrow 6^+_1$ 	& K & 0.025(7) 				& 0.01555(22) 		& 0.0573(8) 	& M1+E2(+E0)\\
\end{tabular}
\end{ruledtabular}
\footnotetext[1]{Obtained indirectly, by comparing the intensity of the $\gamma$ rays. See text for details.}
\footnotetext[2]{Calculated as a difference between the total internal conversion coefficient and the L- and M+-internal conversion coefficients.}
\footnotetext[3]{Limit given with 95\% credible interval.}
\footnotetext[4]{$\alpha_{th}(E1)$ = 0.00597(9)}
\end{table*}

The level at 1872 keV from our study (see the decay scheme in Supplemental Material \cite{supplemental}) has a 1 keV lower excitation energy compared to Ref. \cite{Deng1995} and has a different de-excitation pattern. Thus, unlike the previous $\beta$-decay study \cite{Rapisarda2017}, we propose that our 1872-keV level and the 1873-keV level from Ref. \cite{Deng1995} are two different states.

The 1450-, 2036-, 2093- and 2309-keV $\gamma$ rays have been placed in the decay scheme based on the energy sum arguments (see Supplemental Material \cite{supplemental}). These $\gamma$ rays were not included in the determination of the energy of the excited states.

The 367-keV $2^+_1 \rightarrow 0^+_1$ and the 608-keV $2^+_3 \rightarrow 0^+_2$ transitions are in a mutual coincidence (see Fig. \ref{fig:184Hg_gg}) indicating the existence of the 9-keV $0^+_2 \rightarrow 2^+_1$ transition. To estimate its total intensity ($I_t(9)$), a similar method as in the case of the 512-keV $\gamma$ ray in $^{182}$Hg was used. The number of counts in the 608-keV peak in the spectrum gated on the 367-keV transition $N_{Rg}(608)$ was compared to the number of counts in the same peak in the $\gamma$-ray singles energy spectrum $N_{Rs}(608)$. The $N_{Rg}(608)$ value was corrected by the $\gamma$-gate detection efficiency $\epsilon_\gamma(367)$, by the factor $\frac{3}{4}$ to include the 
reduction of $\gamma$-detection efficiency in coincidence spectrum due to the fact that one out of four germanium detectors is being used for $\gamma$ gating, and by the ICC of the gating transition $\alpha_{tot}(367)$, leading to the following:
\begin{equation}
\frac{I_t(9)}{I_t(9) + I_t(375)} = \frac{N_{Rg}(608)\frac{1+\alpha_{tot}(367)}{\frac{3}{4}\epsilon_\gamma(367)}}{N_{Rs}(608)} = 0.055(11) \equiv B \mathrm{.}
\label{eq:itot9toIsum375}
\end{equation}
To obtain the ratio of the total intensity of the 9 keV transition to the 375 keV transition, one can write:
\begin{equation}
\frac{I_t(9)}{I_t(375)} = \frac{B}{1 - B} = 0.059(12) \equiv R \mathrm{.}
\end{equation}
Having the ratio $R$ and the mean lifetime of the $0^+_2$ state ($\tau = 0.9(3)$ ns \cite{Cole1976}), we were able to calculate $\rho^2(E0;0^+_2 \rightarrow 0^+_1) = 4.1(14) \times 10^{-3}$, as well as $B(E2;0^+_2 \rightarrow 2^+_1) = 0.49(22)~ e^2b^2$. The latter is in $2\sigma$ agreement with $1.3^{+0.7}_{-0.5}$ $e^2b^2$ from the Coulomb excitation studies \cite{Wrzosek-Lipska2019}. 
 
The weak 119.2-keV $4^+_1 \rightarrow 2^+_2$ transition is very close to the strong 119.7-keV $\gamma$-ray originating from the decay of $^{184}$Ir to $^{184}$Os. The $\gamma$-ray intensity of the 119-keV transition $N_\gamma(119;\mathrm{Hg})$ was obtained by subtracting the contribution associated with the osmium line ($N_\gamma(119;\mathrm{Os})$) from the total number of counts in the peak ($N_\gamma(119)$). This contribution was calculated by scaling the number of counts in the strongest osmium peak at 264 keV ($N_\gamma(264;\mathrm{Os})$) by the intensity ratios from Ref. \cite{Kibedi1994}:

\begin{equation}
\frac{N_\gamma(119;\mathrm{Hg})}{\epsilon_\gamma(119)} = \frac{N_\gamma(119)}{\epsilon_\gamma(119)} - \frac{N_\gamma(264;\mathrm{Os})}{\epsilon_\gamma(264)}\frac{I_\gamma(119;\mathrm{Os})}{I_\gamma(264;\mathrm{Os})}  \mathrm{.}
\end{equation}

\noindent By comparing the extracted value with the number of counts in the 287-keV $4^+_1 \rightarrow 2^+_1$ transition, an upper limit of the branching ratio for the 119-keV transition equal to $0.6$ was obtained. The energy of this transition was calculated as the energy difference between the excited states.

A number of $\gamma$ lines could be identified as doublet structures. There are two transitions with an energy around 1179 keV. The intensity of the $1445\mathrm{~keV} \rightarrow 367\mathrm{~keV}$ transition was determined from the $\gamma$-$\gamma$ coincidences, while for the $1179\mathrm{~keV} \rightarrow 0\mathrm{~keV}$ transition it was determined as a difference between the intensity from the $\gamma$-ray singles energy spectrum and the intensity obtained from the coincidence data. The same method was also applied for pairs of transitions at 765 keV ($1854\mathrm{~keV} \rightarrow 1089\mathrm{~keV}$ from coincidence data, $1300\mathrm{~keV} \rightarrow 535\mathrm{~keV}$ as a difference) and at 1082 keV ($2495\mathrm{~keV} \rightarrow 1413\mathrm{~keV}$ from coincidences, $1450\mathrm{~keV} \rightarrow 367\mathrm{~keV}$ as a difference). In addition, the energy of the $1450\mathrm{~keV} \rightarrow 367\mathrm{~keV}$ $\gamma$ ray was determined as the energy difference between the excited states.

\begin{figure}
\includegraphics[width=\columnwidth]{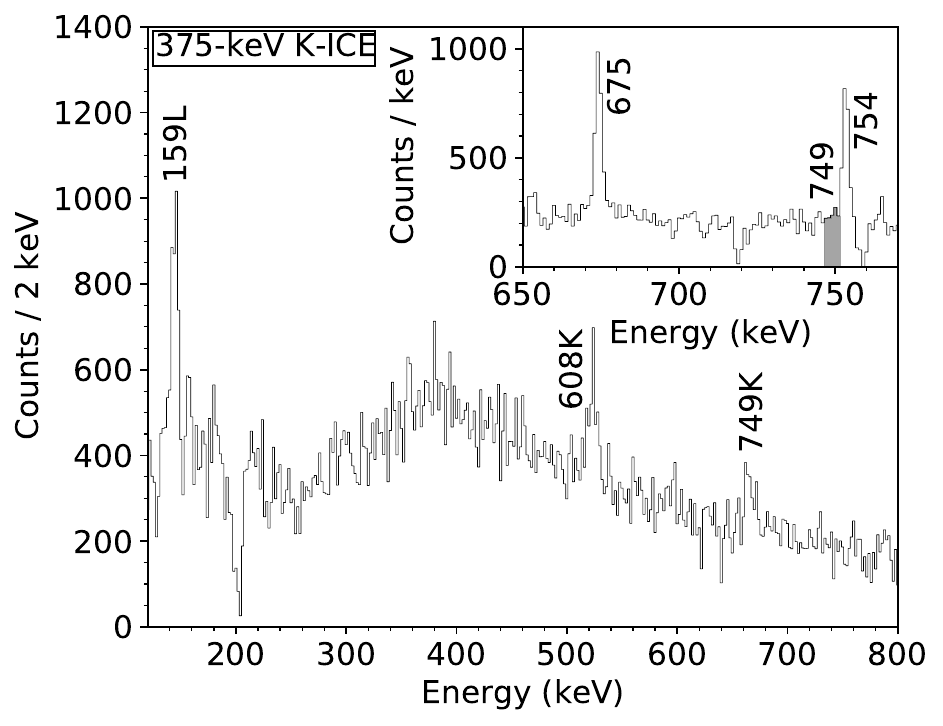}
\caption{\label{fig:ee}Portion of the electron energy spectrum gated on the K-ICE of the 375-keV $0^+_2 \rightarrow 0^+_1$ transition in $^{184}$Hg. The observed peaks are labeled by the energy of the transition they are originate from. Inset: portion of the $\gamma$-ray energy spectrum gated on the K-ICE of the 375-keV $0^+_2 \rightarrow 0^+_1$ transition with the position of the non-observed 749-keV $\gamma$ ray indicated with a shaded area.}
\end{figure}

Based on the electron-electron coincidences gated on the 375-keV $0^+_2\rightarrow 0^+_1$ transition (Fig. \ref{fig:ee}), a state at 1124 keV was identified. The lack of corresponding $\gamma$ ray ($\alpha_K > 1.256$) indicates a strong $E0$ component in the 749-keV transition (Table \ref{tab:conversion184Hg}) and, thus, spin and parity of $0^+$ are attributed to the state. To determine the 749-keV transition's branching ratio, the number of K-ICE in the electron-electron spectrum was compared to the 589-keV $\gamma$-ray transition intensity after correcting them by detection efficiencies as well as a factor $\frac{\Omega_{tot}(749)}{\Omega_K(749)} = 1.2$ to include ICEs from other atomic shells.

The observation of the $0^+_3 \rightarrow 0^+_1$ transition was beyond the observational limit. In addition, there is no known transition feeding the 1124-keV state, thus, an upper limit could not be deduced.

\begin{figure}
\includegraphics[width=\columnwidth]{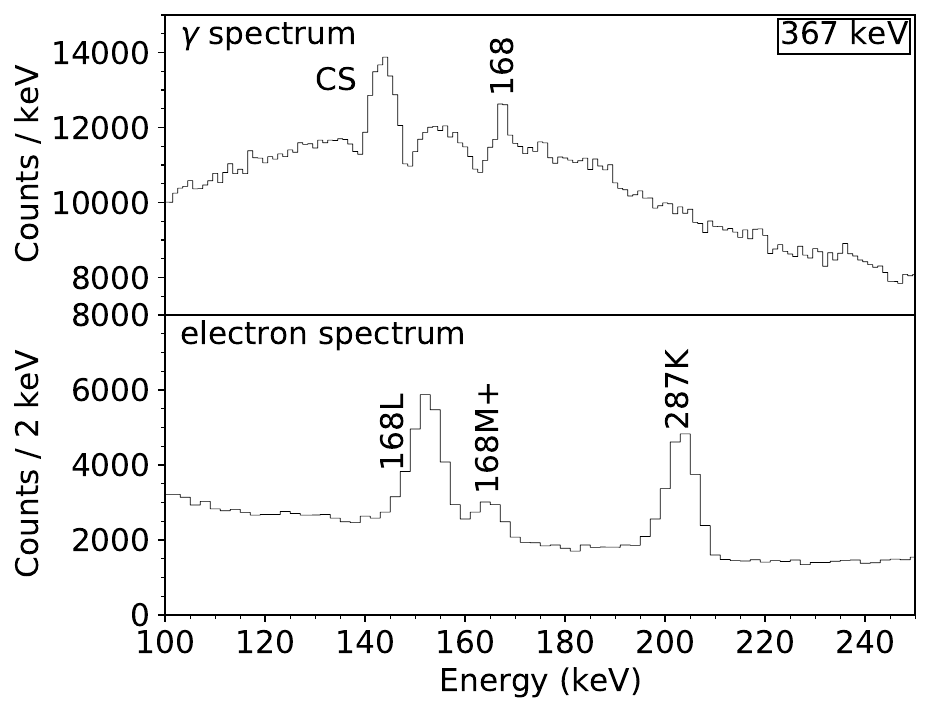}
\caption{\label{fig:168keV}$\gamma$-ray (top) and electron (bottom) energy spectra gated on the 367-keV ($2^+_1 \rightarrow 0^+_1$) $\gamma$ ray in $^{184}$Hg. A peak labeled as `CS' stems for the Compton scattering of strong $\gamma$ rays between different clovers.}
\end{figure}

\begin{figure}
\includegraphics[width=\columnwidth]{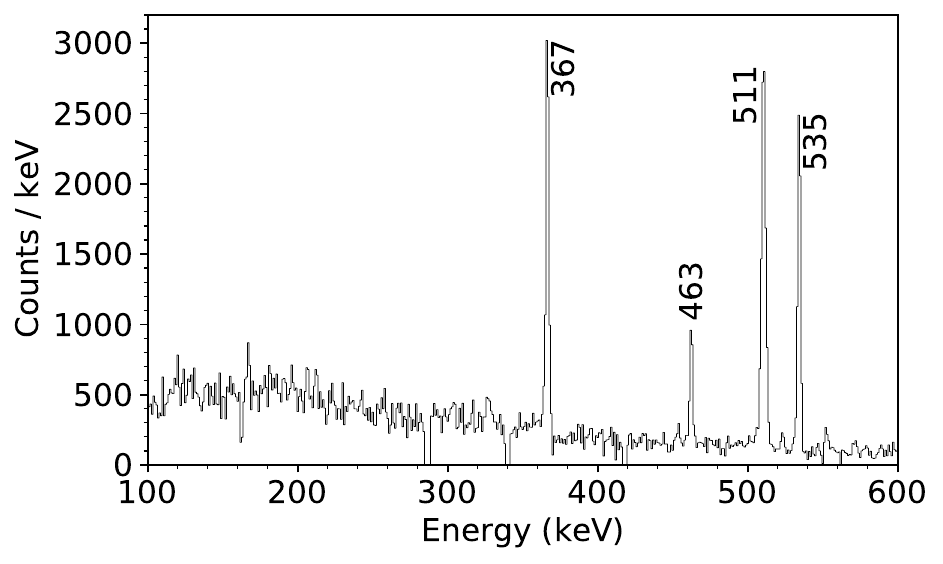}
\caption{\label{fig:imbalance}Portion of the $\gamma$-ray energy spectrum gated on the $\gamma$ rays feeding the 535-keV state in $^{184}$Hg, used to determine the total ICC of the 168-keV $2^+_2 \rightarrow 2^+_1$ transition by the imbalance method, see text for details. Main peaks are labeled by the energy given in keV.}
\end{figure}

The L- and M+-ICCs for the 168-keV $2^+_2 \rightarrow 2^+_1$ transition were obtained from the $\gamma$-ray and electron energy spectra gated on the 367-keV $2^+_1 \rightarrow 0^+_1$ transition, see Fig. \ref{fig:168keV}, whereas the K-ICE energy was below the detection threshold. However, by using the $\gamma$-imbalance method proposed in Ref. \cite{Rapisarda2017}, the total ICC ($\alpha_{tot}(168)$) was extracted by comparing the number of 367- and 535-keV $\gamma$ rays ($I_\gamma(367)$ and $I_\gamma(535)$, respectively) in the $\gamma$-ray energy spectrum gated on the transitions feeding the 535-keV $2^+_2$ state (552, 589, 979, 1022, 1068, 1171, 1270, 1328, 1537, 1548, 1809 and 1883 keV, see Fig. \ref{fig:imbalance}). The ICC can be extracted as follows:
\begin{equation}
\begin{split}
\alpha_{tot}(168) & = \frac{I_\gamma(367) (1 + \alpha_{tot}(367))}{I_\gamma(535) Br_\gamma(168)} \\& - \frac{Br_\gamma(159)}{Br_\gamma(168)}\frac{I_{tot}(9)}{I_{tot}(9) + I_{tot}(375)}(1+\alpha_{tot}(159)) \\& - 1 \mathrm{,}
\end{split}
\end{equation}
where $\alpha_{tot}(367)$ and $\alpha_{tot}(159)$ are the total ICCs of the 367- and 159-keV transitions, respectively, calculated using BrIcc \cite{Kibedi2008}, $Br_\gamma(168)$ and $Br_\gamma(159)$ are the $\gamma$-branching ratios from this analysis (see Tab. III in Supplemental Material \cite{supplemental}) while $\frac{I_{tot}(9)}{I_{tot}(9) + I_{tot}(375)}$ is the intensity ratio of the 9-keV transition, extracted in this work, see Eq. \ref{eq:itot9toIsum375}. The K-ICC was determined as a difference between the total and the L and M+ ICCs. The value obtained in our work ($\alpha_{tot} = 12.8(24)$) is in good agreement with 14.2(36) reported in Ref. \cite{Rapisarda2017}. It should be noted that the main source of uncertainty comes from the precision of the $Br_\gamma(168)$ branching ratio.

From the same gate on the 367-keV $2^+_1 \rightarrow 0^+_1$ transition, the K-ICC of the 617-keV $2^+_3 \rightarrow 2^+_1$ transition was determined. The extracted value $\alpha_{K} = 0.066(6)$ indicates the existence of an $E0$ component and allows us to confirm the spin and parity of $2^+$ for the 984-keV state proposed in the previous work \cite{Rapisarda2017}.

\begin{figure}
\includegraphics[width=\columnwidth]{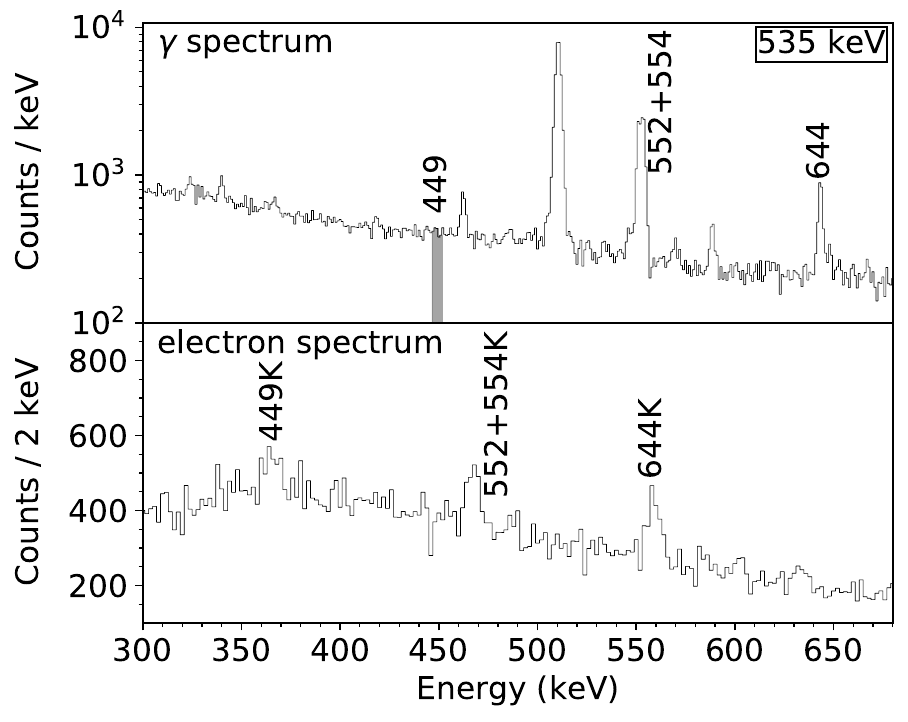}
\caption{\label{fig:535keV}$\gamma$-ray (top) and electron (bottom) energy spectra gated on the 535-keV ($2^+_2 \rightarrow 0^+_1$) $\gamma$-ray in $^{184}$Hg. The main $\gamma$-ray peaks are labeled with the energy given in keV. The position of the non-observed 449-keV $\gamma$ ray is indicated with a shaded area.}
\end{figure}

The K-ICC of the 644-keV $2^+_4 \rightarrow 2^+_2$ transition was obtained from the spectra gated on the 535-keV $\gamma$ ray, see Fig. \ref{fig:535keV}. The extracted value $\alpha_{K} = 0.100(14)$ indicates the existence of an $E0$ component which allows us to confirm the $2^+$ assignment of the 1179-keV state proposed in Ref. \cite{Rapisarda2017}. From the same gate, the lower limit for the K-ICC of the 449-keV transition was extracted ($\alpha_{K} > 0.355$) and the result supports the $2^+$ assignment of the 984-keV level. 

The ICEs from the 552- ($4^+_2 \rightarrow 2^+_2$) and 554-keV ($(3)^+_1 \rightarrow 2^+_2$) transitions create one unresolved peak at 470 keV in the electron energy spectrum, as presented in Fig. \ref{fig:535keV}. In order to obtain the ICC of the 554-keV $\gamma$ ray, deconvolution of the electron peak was needed. The expected number of electrons from the 552-keV $E2$ transition, calculated based on the number of registered $\gamma$ rays, was subtracted from the total number of electrons in the peak. The extracted value is in agreement with a pure $E2$ multipolarity with a possible small admixture of an $M1$ component. This result allows us to propose a positive parity for the 1089-keV state and to keep the previously proposed spin (3) \cite{Rapisarda2017}.
 
The K-ICC of the 646-keV transition was determined from the spectra gated on the yrast 287-keV $4^+ \rightarrow 2^+$ $\gamma$ ray. Although the obtained value, $\alpha_K = 0.072(13)$, has a relatively large uncertainty, it is more than 2$\sigma$ larger than the coefficient of a pure $M1$ transition ($\alpha_K(M1) = 0.0386(6)$), which indicates the existence of an $E0$ component. As a result, we were able to firmly establish the spin and parity of $4^+$ for the 1300-keV state.  

\begin{figure}
\includegraphics[width=\columnwidth]{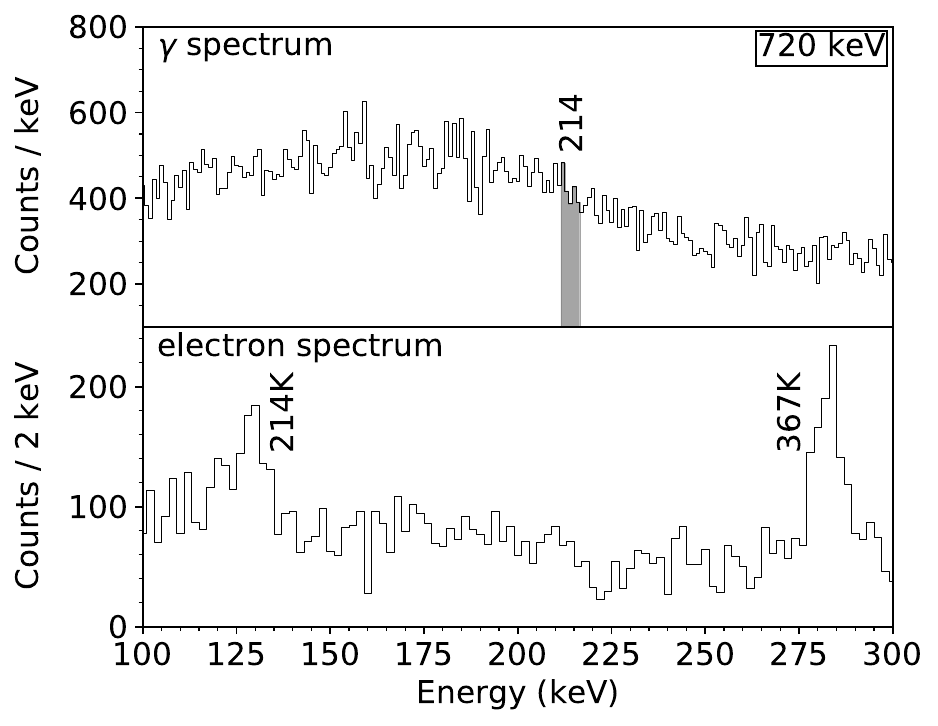}
\caption{\label{fig:214K}$\gamma$-ray (top) and electron (bottom) energy spectra gated on the 720-keV ($4^+_2 \rightarrow 2^+_1$) $\gamma$ ray in $^{184}$Hg. A position of the non-observed 214-keV $\gamma$ ray is indicated with a shaded area.}
\end{figure}

The 214-keV transition de-exciting the 1300-keV state has been observed solely via ICEs (see Fig. \ref{fig:214K}). The limit for the K-ICC (see Tab. \ref{tab:conversion184Hg}), which was extracted from the spectra gated on the 720-keV $\gamma$ rays, implies an $E0$ transition. This conclusion also confirms our $4^+$ assignment to the 1300-keV state.

The ICC of the 556-keV $6^+_2 \rightarrow 6^+_1$ transition was obtained from the $\gamma$-ray and electron energy spectra gated on the 340-keV $\gamma$ ray and points out to a mixed $E2/M1$ multipolarity. However, as in the case of the 586-keV transition in $^{182}$Hg, the existence of an $E0$ component cannot be excluded without an independent measurement of the $\delta$ mixing ratio.

\subsection{\label{sec:186Hg}Excited states in $^{186}$Hg}

\begin{table*}
\caption{\label{tab:conversion186Hg}
Experimental internal conversion coefficients $\alpha_{exp}$ of transitions in $^{182}$Hg compared with the theoretical values $\alpha_{th}$ calculated using BrIcc \cite{Kibedi2008} and the deduced transition multipolarities $X\lambda$.}
\begin{ruledtabular}
\begin{tabular}{ccccccccc}
$E_{i}$ (keV) 		& $E_{f}$ (keV)	& $E_t$ (keV)	& Transition 					& Shell & $\alpha_{exp}$ 			& $\alpha_{th}(E2)$ 	& $\alpha_{th}(M1)$ & $X\lambda$  \\\hline
621.4(1) 			& 405.5(1) 		& 216.0(1) 			& $2^+_2 \rightarrow 2^+_1$ 	& K & 3.5(3)  	& 0.1406(20) 			& 0.732(11) 	& E0+M1+E2\\
 				&  				&  					&  							& L & 0.66(6) 	& 0.1203(17) 			& 0.1229(18) 	& \\ 
 				&  				&  					&  							& M+ & 0.194(18) & 0.0400(5) 			& 0.0372(4) 	& \\ 
1080.8(1) 		& 808.4(1) 		& 272.5(1) 			& $4^+_2 \rightarrow 4^+_1$ 	& K & 0.72(22)	& 0.0796(12) 			& 0.385(6) 		& E0+M1+E2 \\
1434.2(1) 		& 1080.8(1) 		& 353.4(2) 			& $4^+_4 \rightarrow 4^+_2$ 	& K & $>$ 1.54\footnotemark[1]  		& 0.0427(6) 			& 0.190(3) 		& E0(+M1+E2)\\ 
1434.2(1) 		& 808.4(1) 		& 625.9(1) 			& $4^+_4 \rightarrow 4^+_1$ 	& K & 0.022(4) 	& 0.01214(17) 		& 0.0420(6) 	& M1+E2(+E0) \\ 
1678.2(1) 		& 1080.8(1) 		& 597.4(1) 			& $6^+_2 \rightarrow 4^+_2$ 	& K & 0.013(4) 	& 0.01336(19) 		& 0.0474(7) 	& E2 \\
2218.4(1) 		& 1976.3(1) 		& 242.1(1) 			& $(8^-_1) \rightarrow 8^+_2$ 		& K & $<$ 0.055\footnotemark[1]\footnotemark[2]  		& 0.1064(15) 			& 0.533(8) 		& E1 \\
\end{tabular}
\end{ruledtabular}
\footnotetext[1]{Limit given with 95\% credible interval.}
\footnotetext[2]{$\alpha_{th}(E1)$ = 0.0380(6)}
\end{table*}

\begin{figure}
\includegraphics[width=\columnwidth]{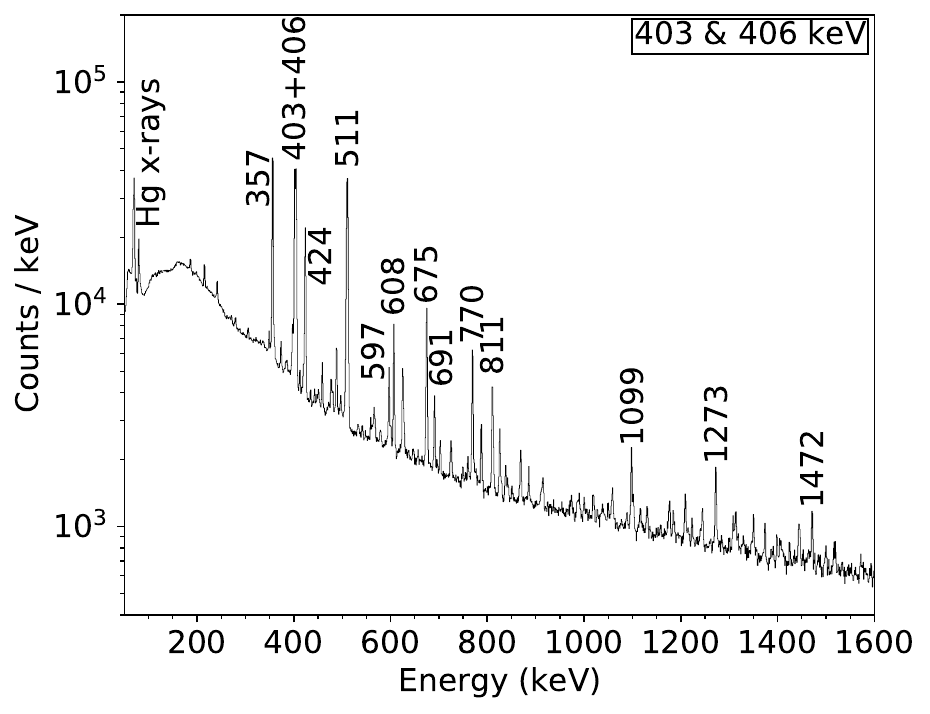}
\caption{\label{fig:186Hg_gg}Portion of the $\gamma$-ray energy spectrum gated on the 403-keV ($4^+_1 \rightarrow 2^+_1$) and 406-keV ($2^+_1 \rightarrow 0^+_1$) $\gamma$-ray transitions in $^{186}$Hg. The most prominent lines have been labeled with the transition energy in keV.}
\end{figure}

\begin{figure}
\includegraphics[width=\columnwidth]{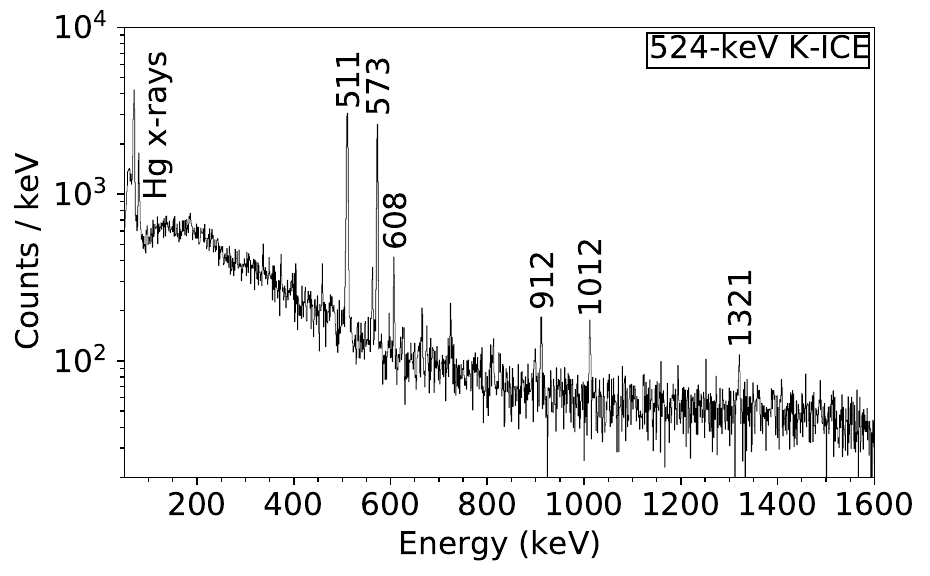}
\caption{\label{fig:186Hg_ge}Portion of the $\gamma$-ray energy spectrum gated on the 524-keV ($0^+_2 \rightarrow 0^+_1$) K-ICE in $^{186}$Hg. Main coincident lines are labeled by the energy given in keV.}
\end{figure}

\begin{figure*}
\includegraphics[width=0.8\textwidth]{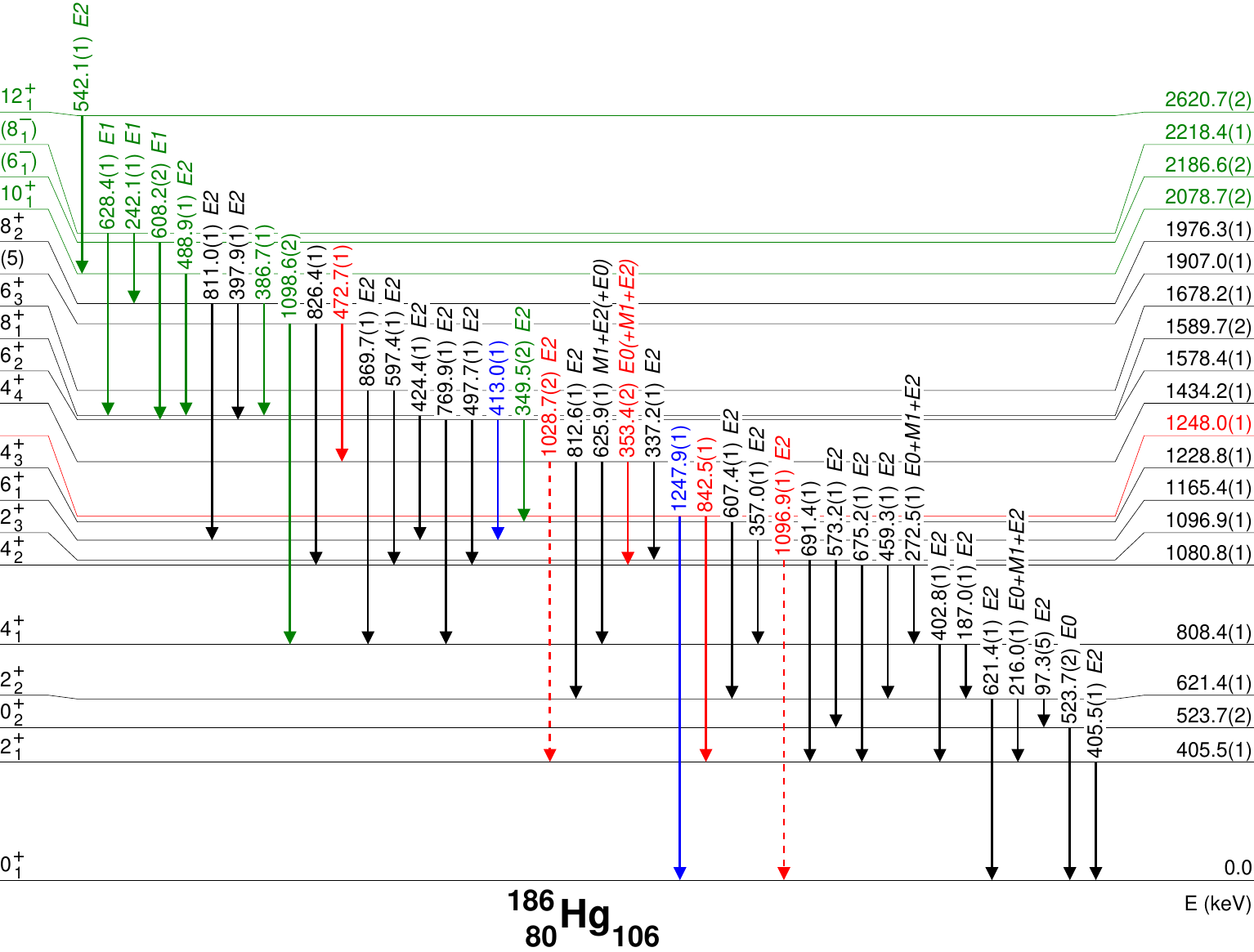}
\caption{\label{fig:186Hgdecayscheme}Partial level scheme of excited states in $^{186}$Hg populated in the $\beta$ decay of $^{186}$Tl extracted in this work. For the full version, see Supplemental Material \cite{supplemental}. Levels and transitions known from the previous $\beta$-decay studies are plotted in black, shifted in the decay scheme in blue, known from other studies in green and newly identified in red. Transitions not observed in this work for which the intensity limits have been determined are plotted with dashed lines. Spins, parities and proposed transition multipolarities are taken from this work and Ref. \cite{Batchelder2022}.}
\end{figure*}

Based on the coincidence analysis, we confirmed most of the decay scheme reported in the latest evaluation \cite{Batchelder2022} and substantially extended it. Typical spectra are presented in Figs. \ref{fig:186Hg_gg} and \ref{fig:186Hg_ge} while portions of the $\gamma$-ray and electron singles energy spectra are presented in Fig. 7 in Ref. \cite{Stryjczyk2020}. In total, 102 excited states and 156 transitions were associated with $^{186}$Hg, including 91 new transitions and 68 new levels. Nine states and 17 transitions known from the in-beam studies \cite{Batchelder2022} have been also observed in this $\beta$-decay study. The summary of the measured $\gamma$ rays with the branching ratios is presented in Tables V and VI in Supplemental Material \cite{supplemental} and the extracted ICCs are summarized in Table \ref{tab:conversion186Hg}. The partial decay scheme is presented in Fig. \ref{fig:186Hgdecayscheme}.

\begin{figure}
\includegraphics[width=\columnwidth]{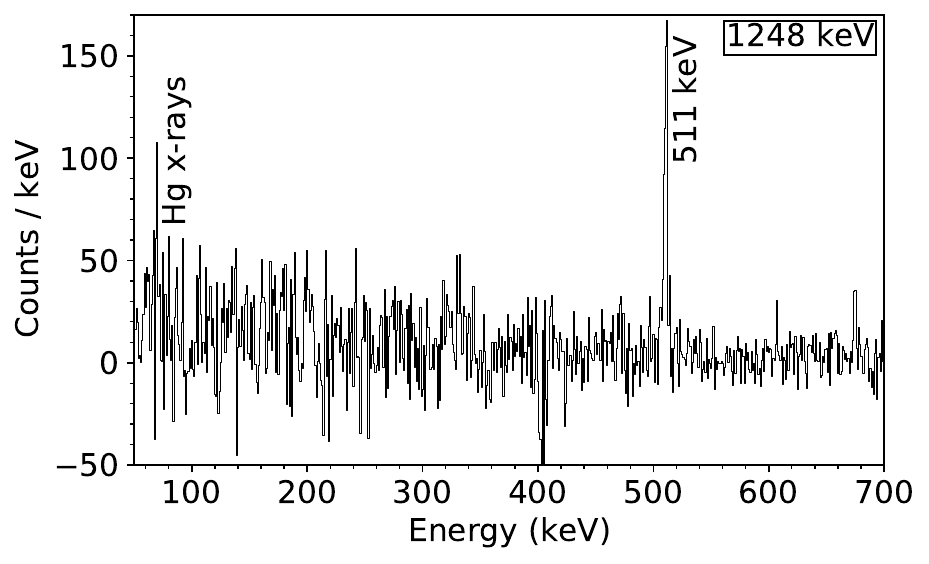}
\caption{\label{fig:1247keV}Portion of the $\gamma$-ray energy spectrum gated on the 1248-keV $\gamma$ ray in $^{186}$Hg.}
\end{figure}

Compared to the previous $\beta$-decay studies \cite{Beraud1977,Batchelder2022}, three previously unplaced transitions, 413, 726 and 1273 keV, were put in the decay scheme based on the $\gamma$-$\gamma$ coincidence data. It should be noted that the placement of the 413-keV $\gamma$ ray is in agreement with the in-beam studies \cite{Janssens1984}. We were not able to confirm the existence of two excited states at 1966 and 2056 keV, which are reported in the evaluation \cite{Batchelder2022}. The former was supposed to de-excite via the emission of a 288-keV $\gamma$ ray, which has not been observed, while the latter was proposed to decay by emitting a 1248-keV $\gamma$ ray. In our analysis, this transition is in coincidence only with the Hg x rays and the 511-keV annihilation peak (see Fig. \ref{fig:1247keV}). Based on these coincidences and the fact it has the same energy as the new 1248-keV state established in the $\gamma$-$\gamma$ coincidence analysis, we propose it de-excites this level to the ground state.

\begin{figure}[h!t!b]
\includegraphics[width=\columnwidth]{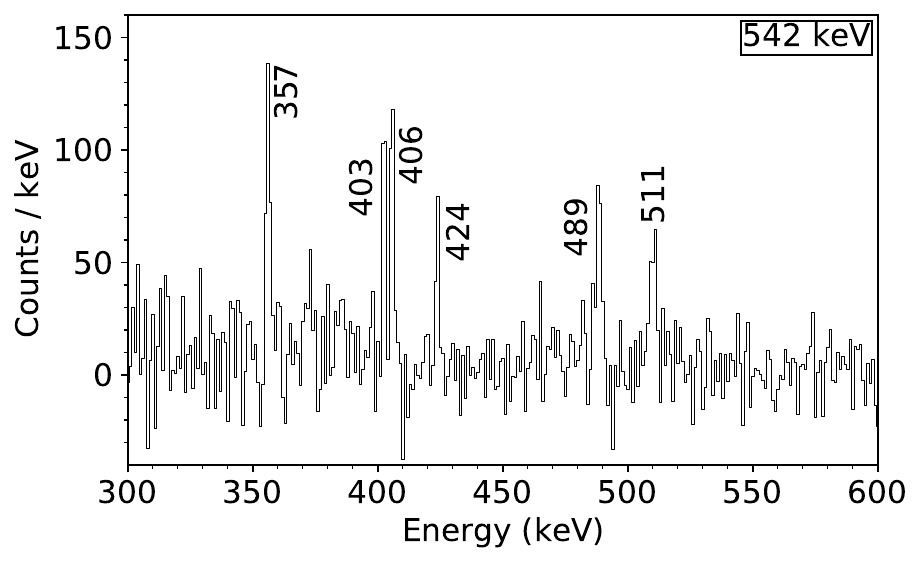}
\caption{\label{fig:12pto10p}Portion of the $\gamma$-ray energy spectrum gated on the 542-keV $12^+_1 \rightarrow 10^+_1$ $\gamma$ ray in $^{186}$Hg. The yrast transitions and the 511-keV annihilation peak are labeled by the energy given in keV.}
\end{figure}

The observation of the de-excitation of the 2621-keV $12^+_1$ state (see Fig. \ref{fig:12pto10p}) points out to a $\beta$-decay branch of the $^{186}$Tl $10^{(-)}$ isomeric state. The $^{186}\mathrm{Tl}(7^{(+)}) \xrightarrow{\beta^+/EC}\mathrm{}^{186}\mathrm{Hg}(12^+)$ decay would have a fourth-forbidden unique character while the $^{186}\mathrm{Tl}(10^{(-)}) \xrightarrow{\beta^+/EC}\mathrm{}^{186}\mathrm{Hg}(12^+)$ would be much more probable first-forbidden unique decay. More information regarding the decay of the $^{186}$Tl $10^{(-)}$ isomeric state can be found in Ref. \cite{Stryjczyk2020}.

\begin{figure}
\includegraphics[width=\columnwidth]{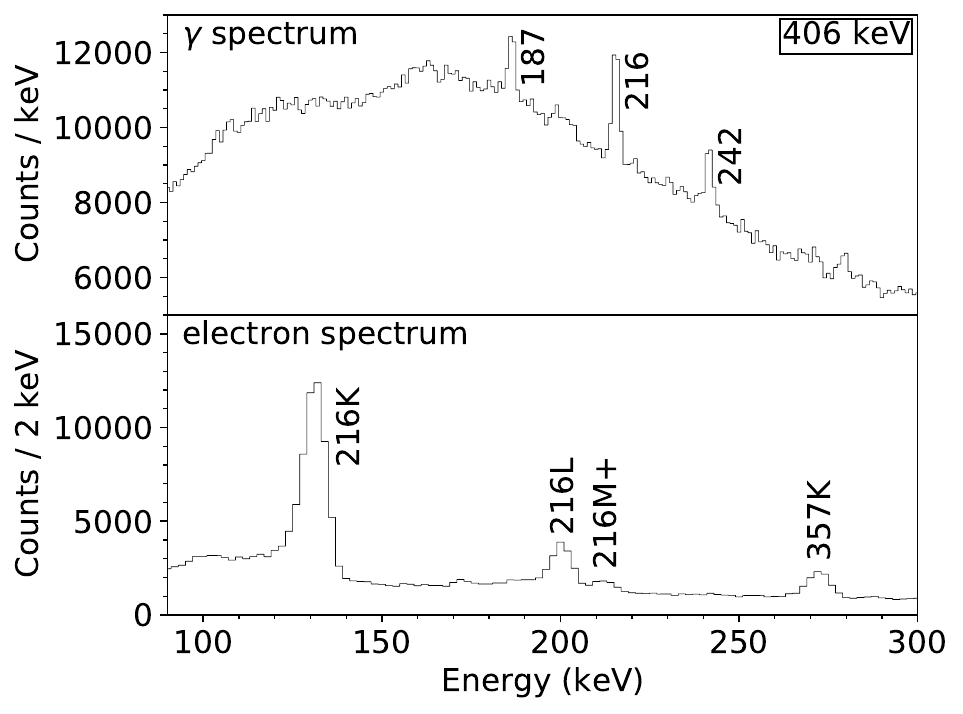}
\caption{\label{fig:216K}$\gamma$-ray (top) and electron (bottom) energy spectra gated on the 406-keV ($2^+_1 \rightarrow 0^+_1$) $\gamma$ ray in $^{186}$Hg.}
\end{figure}

The ICCs of the 216-keV $2^+_2 \rightarrow 2^+_1$ transition were obtained from the spectra gated on the yrast 406-keV $2^+_1 \rightarrow 0^+_1$ $\gamma$ ray (see Fig. \ref{fig:216K}). The values (see Table \ref{tab:conversion186Hg}) are lower than the $\alpha_K = 4.9(13)$ and $\alpha_L = 1.03(26)$ values reported in Ref. \cite{Scheck2011} but in agreement within 2$\sigma$. 

\begin{figure}
\includegraphics[width=\columnwidth]{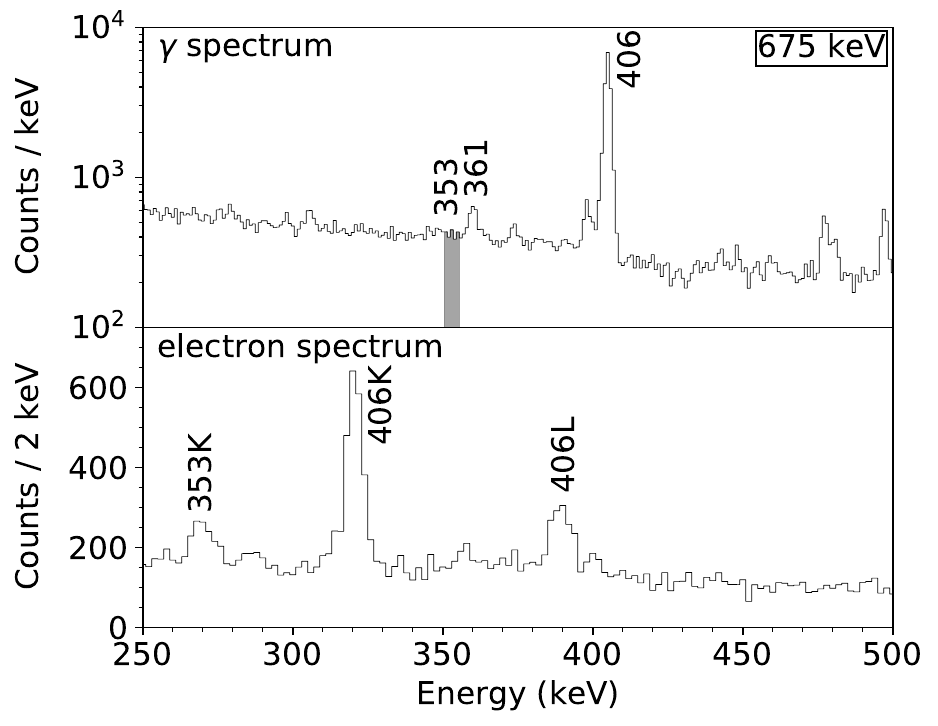}
\caption{\label{fig:353K}$\gamma$-ray (top) and electron (bottom) energy spectra gated on the 675-keV ($4^+_2 \rightarrow 2^+_1$) $\gamma$ ray in $^{186}$Hg. The position of the non-observed 353-keV $\gamma$ ray is marked with a shaded area.}
\end{figure}

The 353-keV transition de-exciting the state at 1434 keV was observed only via ICEs, see Fig. \ref{fig:353K}. The limit for the K-ICC ($\alpha_K > 1.54$) was extracted from the spectra gated on the 675-keV $4^+_2 \rightarrow 2^+_1$ $\gamma$ ray and it points out to a presence of a strong $E0$ component. As a result, the previously proposed ($3^+$) assignment of the 1434-keV state \cite{Delaroche1994} was changed to $4^+$. By employing the same $\gamma$-ray energy gate at the 675-keV transition, the K-ICC of the 597-keV $6^+_2 \rightarrow 4^+_2$ transition was extracted and the $E2$ multipolarity of this transition was confirmed. It should be noted that the spin-parity assignments reported in the ENSDF evaluation \cite{Batchelder2022} for the 1660.0-, 1868.9-, 2138.8- and 2428.4-keV states (see Supplemental Material for the full decay scheme \cite{supplemental}) were based on the same theoretical calculations as for the 1434-keV state \cite{Delaroche1994}. Since the assignment was incorrect for one state, we do not adopt them for other levels.

By gating on the 403-keV $4^+_1\rightarrow 2^+_1$ transition, K-ICC of the 272- ($4^+_2\rightarrow 4^+_1$) and 626-keV ($4^+_4\rightarrow 4^+_1$) transitions were extracted. In spite of large uncertainty, related mostly to the limited $\gamma$-ray statistics, it is firmly established that the 272-keV transition has an $E0$ component while in the case of the 626-keV line the value indicates a mixed $E2/M1$ multipolarity. However, similarly to the $6^+_2 \rightarrow 6^+_1$ transition in $^{182,184}$Hg, the $E0$ component cannot be excluded without an independent measurement of the $\delta$ mixing ratio.

\begin{figure}
\includegraphics[width=\columnwidth]{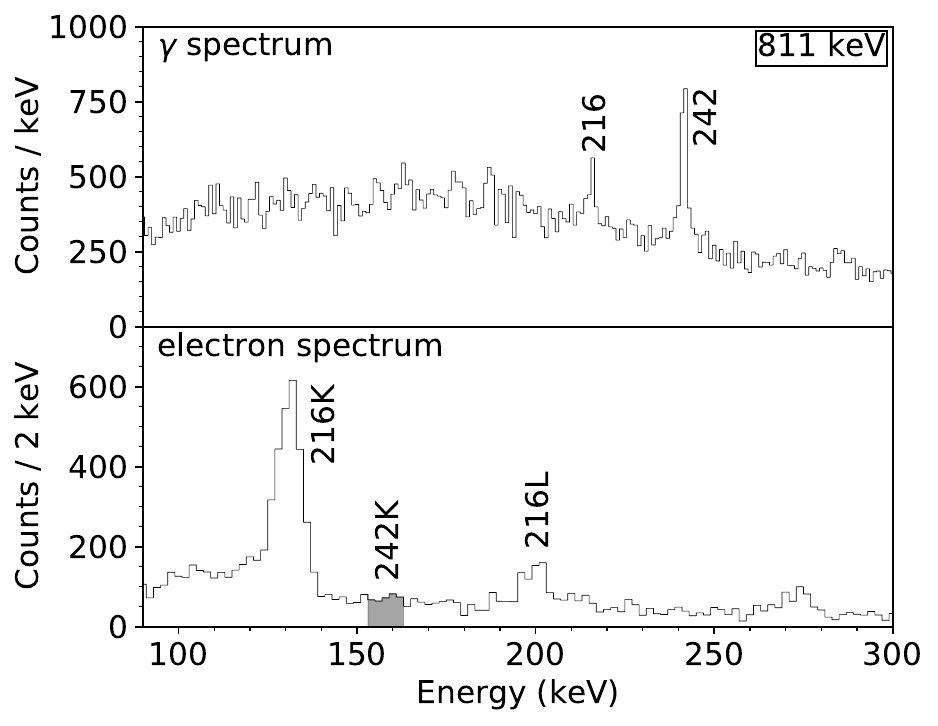}
\caption{\label{fig:242K}$\gamma$-ray (top) and electron (bottom) energy spectra gated on the 811-keV ($8^+_2 \rightarrow 6^+_1$) $\gamma$ ray in $^{186}$Hg. The position of the non-observed K-ICE from the 242-keV transition is indicated with a shaded area.}
\end{figure}

The upper limit for the K-ICC of the 242-keV ${(8^-_1)\rightarrow 8^+_2}$ transition de-exciting the ($8^-_1$) K isomer ($T_{1/2} = 82(5)$ $\mu s$ \cite{Batchelder2022}) was obtained from the spectra gated on the 811-keV $\gamma$ ray (see Fig. \ref{fig:242K}). The result allows us to firmly establish an $E1$ multipolarity and leads to a positive parity assignment for the 1976-keV state. Since this state belongs to the band built on top of the 1229-keV state \cite{Porquet1992,Ma1993,Baglin2003}, we propose a positive parity for all the band members. This result resolves a discrepancy regarding the spin and parity of the 1229-keV state, pointed out in the previous ENSDF evaluation \cite{Baglin2003}, and is in agreement with the $4^+$ assignment proposed in the most recent evaluation \cite{Batchelder2022}. 

\subsection{\label{sec:delta}Multipole mixing ratios}

The determination of K-, L- and M+-ICCs for the $2^+_2 \rightarrow 2^+_1$ transitions in all three isotopes allowed us to determine the $q_K^2(E0/E2)$ and $\delta(E2/M1)$ mixing ratios \cite{Dowie2020}. The experimental ICC of the $E0+M1+E2$ transition from the $i$ atomic shell ($i = K, L, ...$) can be expressed as \cite{Dowie2020}: 

\begin{equation}
\alpha_i^{exp} = \frac{\alpha_i(M1) + \delta^2(1+q_i^2)\alpha_i(E2)}{1+\delta^2} \mathrm{,}
\end{equation}

\noindent where $\alpha_i(M1)$ and $\alpha_i(E2)$ are the calculated ICCs for pure $M1$ and $E2$ transitions, respectively, while $\delta^2$ and $q_i^2$ are the aforementioned mixing ratios. 

The $q^2_i$ values for different atomic shells $i$ and $j$ are linked with the following relation \cite{Dowie2020}:
\begin{equation}
q_j^2 = q_i^2 \times \frac{\Omega_j(E0)}{\Omega_i(E0)} \times \frac{\alpha_i(E2)}{\alpha_j(E2)} \mathrm{,}
\end{equation}
where $\Omega_{i}(E0)$, $\Omega_{j}(E0)$ are the theoretical electronic factors for $E0$ transitions.

\begin{figure}
\includegraphics[width=\columnwidth]{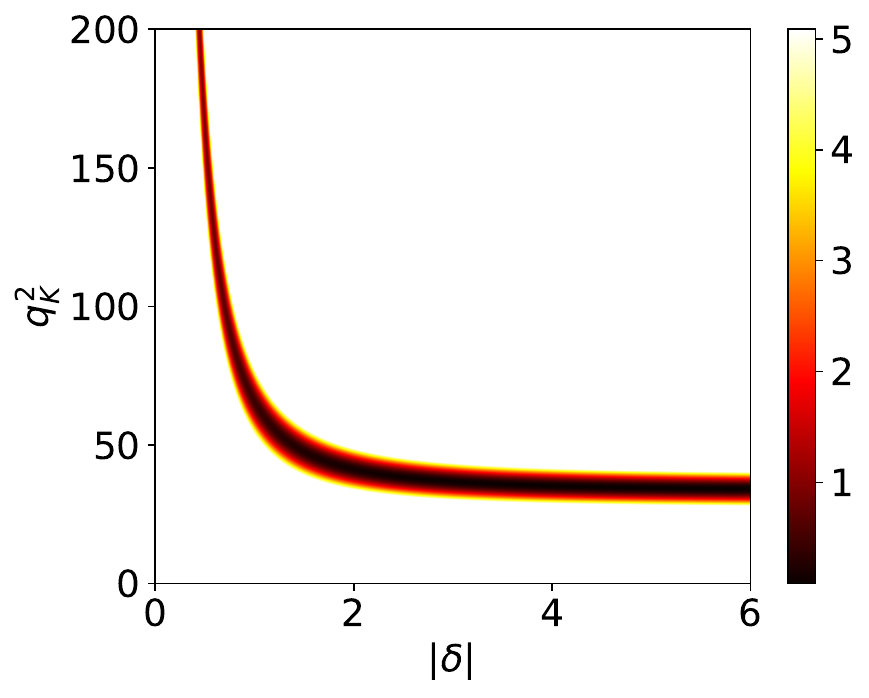}
\caption{\label{fig:deltaqmixing182Hg}The $\chi^2$ plot as a function of the $|\delta|$ and $q_K^2$ mixing parameters of the $2^+_2 \rightarrow 2^+_1$ 197 keV transition in $^{182}$Hg. The $\chi^2$ value is given by the color scale.}
\end{figure}

By having two or more ICCs, the likelihood function $\chi^2$ can be written as:

\begin{equation}
\chi^2 = \sum_i \frac{(\alpha_i^{exp} - \frac{\alpha_i(M1) + \delta^2(1+q_i^2)\alpha_i(E2)}{1+\delta^2})^2}{s_{\alpha_i^{exp}}^2}\mathrm{,}
\end{equation}

\noindent where $s_{\alpha_i^{exp}}$ being the uncertainty of the experimental ICC $\alpha_i^{exp}$. Free parameters were restricted to $q_K^2 < 1000$ and $|\delta| < 10$ by setting priors. The posterior density functions (pdf) were obtained using the Markov Chain Monte Carlo method \cite{thesis}. A pdf for $^{182}$Hg is shown in Fig. \ref{fig:deltaqmixing182Hg}. Values reported in Table \ref{tab:mixing} are the medians and 16th and 84th percentiles of the marginalized pdf or, in cases where only limits are provided, the 5th percentiles. 

The extracted $\delta$ mixing ratio limits are in line with $\delta=1.85$ used in Ref. \cite{Wrzosek-Lipska2019} to determine $\rho^2(E0;2^+_2\rightarrow 2^+_1)$ in $^{182,184}$Hg. The $q_K^2$ values from our work and from Ref. \cite{Kibedi2022} are in agreement for $^{184,186}$Hg but not for $^{182}$Hg, where the literature value of $q_K^2 = 28^{+7}_{-8}$ is more than 3$\sigma$ away from our result. This indicate a stronger contribution of the $E0$ component in the $2^+_2\rightarrow 2^+_1$ transition.

The extracted mixing ratios, together with the $2^+_2$ states lifetimes, can be used to reevaluate the $\rho^2(E0;2^+_2\rightarrow 2^+_1)$ values. However, we note that the known lifetimes are extracted from the Coulomb excitation study \cite{Wrzosek-Lipska2019} and they depend on the spectroscopic input from the previous experiments. The new branching ratios and conversion coefficients from this work will lead to a different set of matrix elements in the Coulex analysis and, as a result, different lifetimes and monopole strengths.

\begin{table}
\caption{\label{tab:mixing}
$q_K^2$ and $|\delta|$ mixing ratios for the $2^+_2 \rightarrow 2^+_1$ transitions.}
\begin{ruledtabular}
\begin{tabular}{cccc}
Nucleus 		& $E_\gamma$ (keV) 	& $q_K^2$				& $|\delta|$ \\\hline
$^{182}$Hg	& 196.5 				& 50$^{+33}_{-7}$ 			& $>0.29$\footnotemark[1] 	\\
$^{184}$Hg	& 168.0 				& 50$^{+61}_{-7}$			& $>0.33$\footnotemark[1] 	\\
$^{186}$Hg	& 216.0 				& $>21$\footnotemark[1] 	& $>0.16$\footnotemark[1] 	\\
\end{tabular}
\end{ruledtabular}
\footnotetext[1]{Limit given with 95\% credible interval.}
\end{table}

\section{\label{sec:discussion}Discussion}

\subsection{\label{sec:spins}Spin and parity assignments}

In the previous sections, the spin and parity of a number of states was determined on the basis of the measured ICCs. The analysis of the de-excitation paths allows us to assign spins and parities of several low-lying states. The details are discussed below.

\underline{$^{182}$Hg, 1507 keV, J$^\pi$ = $3^-$,$4^+$:} this state de-excites solely to the $2^+$ states and it is fed from the ($5^-$) state. Since none of the discussed states exhibits isomeric properties, the only considered transition multipolarities are $E1$, $M1$ and $E2$. That leads to two possible spins, $3^-$ and $4^+$. This assignment allows us to propose the same spins for the 1719 keV level, as they are connected by a transition with an $E0$ component.

\underline{$^{182}$Hg, 1531 keV, J$^\pi$ = (6)$^+$:} the ICC between this and the $6^+_1$ states indicates an $E2/M1$ character and, thus, a positive parity, while the decay to the $4^+$ and $6^+$ states and the similar energies of the $6^+_2$ states in $^{180,184}$Hg (1504 keV \cite{Elseviers2011} and 1550 keV, respectively) suggest a tentative spin assignment of (6). Since this state was proposed in Ref. \cite{Bindra1995} to be a bandhead of band 7 (see Fig. 3 of \cite{Bindra1995}), with levels being connected by $E2$ transitions, we propose that the states belonging to this band, including the 1942-keV state observed in our work, have spins and parities from (8)$^+$ to (16)$^+$. 

\underline{$^{182}$Hg, 1547 keV, J$^\pi$ = $4^+$:} this state feeds the yrast $2^+$, $4^+$ and $6^+$ states.

\underline{$^{182}$Hg, 1985, 2037, 2342, 2418 and 2448 keV, J$^\pi$ = ($5^-$):} there are significant differences in the decay pattern of these states in $^{182}$Hg compared to states at similar excitation energy in $^{184,186}$Hg - in the latter nuclei the excited states de-excite by emission of no more than four different $\gamma$ rays while in the former, five or more de-excitation paths exist. All these states feed the $4^+$ and $6^+$ states and do not feed the $2^+$ and $8^+$ states, which indicates spin 5. In addition, in the $\beta$-decay of $^{180}$Tl($4^-$) to $^{180}$Hg \cite{Elseviers2011}, similar states at 1797 and 2348 keV were observed and both of them had low log(\textit{ft}) values, which suggests an allowed decay and, consequently, a negative parity. The measurement of the magnetic dipole moments by means of laser spectroscopy suggested a similarity in the structure of $^{180}$Tl($4^-$) ground state and the low-spin $^{182}$Tl($4^-$) isomer \cite{Barzakh2017}. Although in our study we cannot extract log(\textit{ft}) values, based on the presented arguments we tentatively propose spin ($5^-$) to the 1985, 2037, 2342, 2418 and 2448 keV levels.

\underline{$^{184}$Hg, 1413.3 keV, J$^\pi$ = ($5^+$):} spin (5) was proposed in previous studies \cite{Deng1995,Baglin2010} and the de-excitation to the (3)$^+$ state suggests a positive parity. This assignment allows us to propose spins ($7^+$) and ($9^+$) to 1803- and 2257-keV states, respectively, as they are connected by $E2$ transitions \cite{Deng1995}.
 
\subsection{\label{sec:models}Comparison with the theoretical models}

The experimental results were compared to calculations from two theoretical models available in the literature: Interacting Boson Model with Configuration Mixing (IBM-CM) which employs the D1M parametrization of the Gogny energy density functional (IBM Gogny) \cite{Nomura2013} and the Beyond Mean-Field based model (BMF) which uses the SLy6 parametrization of the Skyrme interaction \cite{Yao2013}. Furthermore, additional calculations have been performed within the IBM-CM approach with the phenomenological parametrization (IBM Phen) \cite{Garcia-Ramos2014}, the General Bohr Hamiltonian (GBH) method \cite{Prochniak2009,Prochniak2015,Wrzosek-Lipska2012,Wrzosek-Lipska2019} as well as the symmetry-conserving configuration mixing (SCCM) model \cite{Rodriguez2010,Rodriguez2014,Siciliano2020}.

\begin{figure*}[h!t!b]
\includegraphics[width=\textwidth]{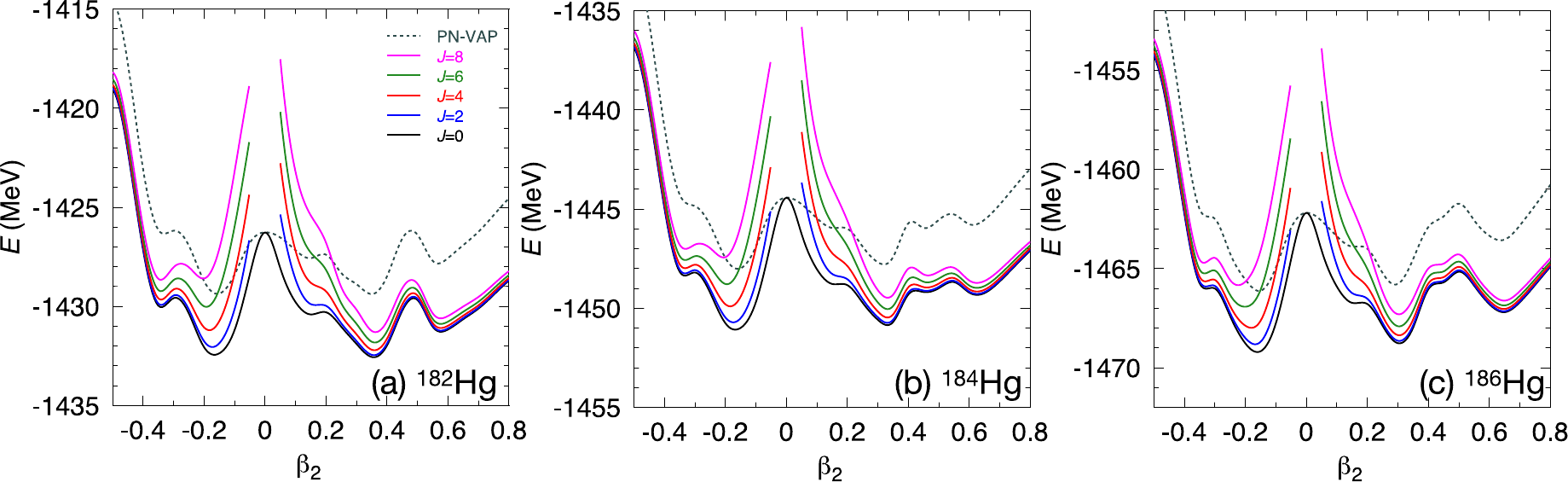}
\caption{\label{fig:sccmprojections}The energy curves as a function of the deformation parameter $\beta_2$ obtained within the SCCM calculations for (a) $^{182}$Hg, (b) $^{184}$Hg and (c) $^{186}$Hg, after particle-number projection (PN-VAP method, dashed lines) and angular momentum projection (PNAMP method, continuous lines).}
\end{figure*}

The first information regarding the structure of $^{182,184,186}$Hg can be retrieved by analyzing the potential energy surfaces (PES) as a function of deformation. In case of SCCM, the curve obtained with the particle-number variation after projection (PN-VAP) method \cite{Anguiano2001} points to a complex structure, with a global oblate minimum at $\beta_2 \approx -0.15$, two normal-deformed (ND) prolate minima at $\beta_2 \approx 0.1$ and 0.25 and one super-deformed (SD) prolate minimum at $\beta_2 \approx 0.6$ (Fig. \ref{fig:sccmprojections}). Furthermore, there is one additional minimum in $^{184}$Hg at $\beta_2 \approx 0.45$ and in $^{182}$Hg at $\beta_2 \approx -0.35$. A projection of PN-VAP wave functions onto angular momentum creates a particle-number and angular momentum projection (PNAMP) set whose structure remains rather unchanged for $J=0$, with the global ND oblate minimum at $\beta_2 \approx -0.17$ and a prolate minimum at $\beta_2 \approx 0.3$ at almost identical energy. One exception is an appearance of a shallow ND oblate minimum at $\beta_2 \approx -0.35$ in both $^{184}$Hg and $^{186}$Hg. These results are consistent with the recent laser spectroscopy study which determined the ground state $|\beta_2|$ value to be about 0.2 \cite{Marsh2018,Sels2019}.

\begin{figure*}
\includegraphics[width=\textwidth]{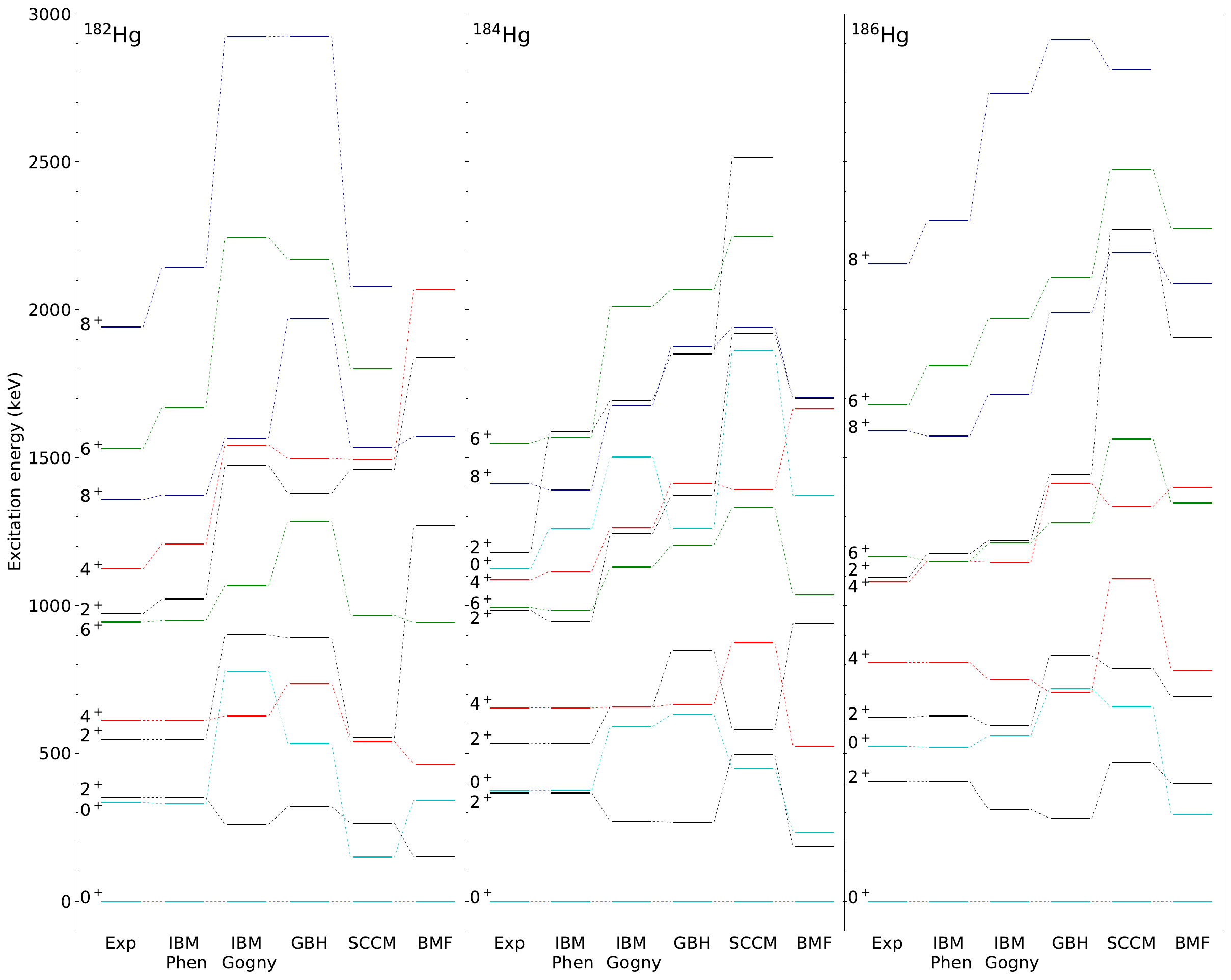}
\caption{\label{fig:expvsthHg}Comparison of the experimental energies of the selected excited states in $^{182,184,186}$Hg with the theoretical calculations. The dashed lines, provided to guide the eye, are connecting the states of the same spin and order.}
\end{figure*}

Comparisons of the experimental energies of excited states with the theoretical predictions are presented in Fig. \ref{fig:expvsthHg}. The best agreement is obtained with IBM Phen but it should be kept in mind that this model was fitted to the experimental data. The only significant discrepancy can be observed for the energy of the $2^+_4$ state in $^{184}$Hg. At the same time, the IBM Gogny calculations reproduce rather poorly the excitation energies with the exception of $^{186}$Hg. It might be related to the fact that for $^{182,184}$Hg these calculations predict strongly deformed ground-state bands and weakly deformed bands built on top of the $0^+_2$ states \cite{Nomura2013} which contradicts the experimental findings \cite{Marsh2018,Sels2019}. On the other hand, for $^{186}$Hg the ground-state band is predicted to be weakly oblate-deformed \cite{Nomura2013}. 

The results from GBH, SCCM and BMF show that the energy differences between the calculated states belonging to the same band are systematically larger than the experimental values, but this is a known deficiency of these calculations \cite{Wrzosek-Lipska2019}. A very poor reproduction of the third $0^+$ and $2^+$ states in SCCM and BMF might be related to the restriction to only axial deformations. 

\begin{figure*}
\includegraphics[width=\textwidth]{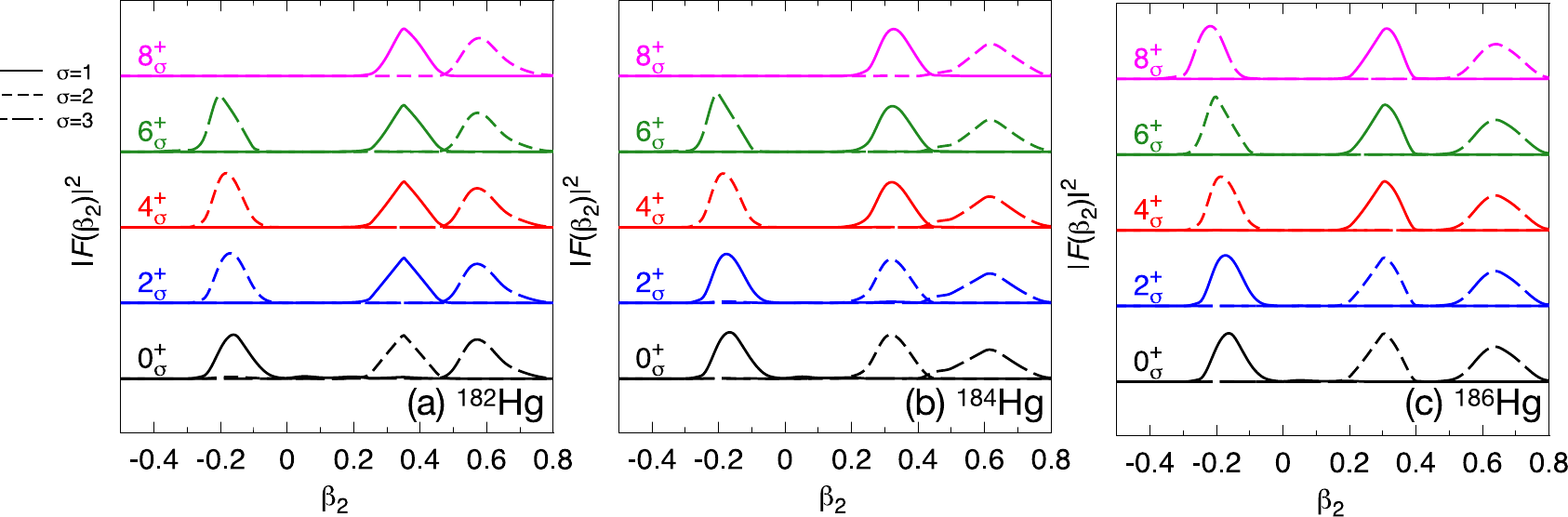}
\caption{\label{fig:sccmwf}Collective wave functions from the SCCM calculations for (a) $^{182}$Hg, (b) $^{184}$Hg and (c) $^{186}$Hg. The first, second and third states of each spin are plotted in full, short-dashed and dashed lines, respectively.}
\end{figure*}

\begin{table}
\caption{\label{tab:models182Hg}
Comparison of the experimental $B(E2)$ ratios in $^{182}$Hg with the theoretical models (see Fig. \ref{fig:182Hgdecayscheme} for the decay scheme). Symbol ``-'' indicates that the particular ratio was not calculated in a given model.}
\begin{ruledtabular}
\begin{tabular}{ccccccc}
Ratio & Exp			& IBM & IBM & GBH 	& SCCM 		& BMF	\\
& & Phen & Gogny & & & \\\hline
$\frac{B(E2;2^+_1\rightarrow 0^+_2)}{B(E2;2^+_1\rightarrow 0^+_1)}$	& 2.9(4) 		& 2.4			& 0.069			& 0.4	& 26 		& 0.072	\\[0.2cm]
$\frac{B(E2;2^+_2\rightarrow 0^+_2)}{B(E2;2^+_2\rightarrow 0^+_1)}$	& 5.5(7) 		& 5.5			& 7.9			& 4.0	& 0.087 		& 787	\\[0.2cm]
$\frac{B(E2;4^+_2\rightarrow 2^+_2)}{B(E2;4^+_2\rightarrow 2^+_1)}$	& 8.5(6) 		& 8.5			& 94			& 103	& 2413 		& 153	\\[0.2cm]
$\frac{B(E2;6^+_2\rightarrow 4^+_2)}{B(E2;6^+_2\rightarrow 4^+_1)}$	& 168(13) 		& 596			& 843			& 949	& 1.12 		& -		\\[0.2cm]
\end{tabular}
\end{ruledtabular}
\end{table}

\begin{table}
\caption{\label{tab:models184Hg}
Comparison of the experimental $B(E2)$ ratios in $^{184}$Hg with the theoretical models (see Fig. \ref{fig:184Hgdecayscheme} for the decay scheme). Symbol ``-'' indicates that the particular ratio was not calculated in a given model.}
\begin{ruledtabular}
\begin{tabular}{ccccccc}
Ratio 																& Exp			& IBM 	& IBM 	& GBH 	& SCCM 		& BMF		\\
& & Phen 	& Gogny & 	& 		& 		\\\hline
$\frac{B(E2;2^+_2\rightarrow 0^+_2)}{B(E2;2^+_2\rightarrow 0^+_1)}$	& 13.5(14) 		& 23.5	& 13.5  & 5.6	& 2298 		& 16.4		\\[0.2cm]
$\frac{B(E2;4^+_1\rightarrow 2^+_2)}{B(E2;4^+_1\rightarrow 2^+_1)}$	& $<0.46$ 		& 0.45	& 0.020  & 0.014	& 32 		& 0.0047 		\\[0.2cm]
$\frac{B(E2;4^+_2\rightarrow 2^+_2)}{B(E2;4^+_2\rightarrow 2^+_1)}$	& 5.5(4) 		& 3.5	& 36.9  & 119	& 0.046 		& $6.4 \times 10^4$	\\[0.2cm]
$\frac{B(E2;2^+_4\rightarrow 0^+_2)}{B(E2;2^+_4\rightarrow 0^+_1)}$	& 3.5(4) 		& 6.3	& 19.8  & 14.4		& - 			& -			\\[0.2cm]
$\frac{B(E2;6^+_2\rightarrow 4^+_2)}{B(E2;6^+_2\rightarrow 4^+_1)}$	& 93(7) 		& 801	& 245  & 811	& 0.048 		& -			\\[0.2cm]
\end{tabular}
\end{ruledtabular}
\end{table}

\begin{table}
\caption{\label{tab:models186Hg}
Comparison of the experimental $B(E2)$ ratios in $^{186}$Hg with the theoretical models (see Fig. \ref{fig:186Hgdecayscheme} for the decay scheme). Symbol ``-'' indicates that the particular ratio was not calculated in a given model.}
\begin{ruledtabular}
\begin{tabular}{ccccccc}
Ratio 																& Exp			& IBM 	& IBM 	& GBH 	& SCCM 		& BMF		\\
& & Phen 	& Gogny & 	& 		& 		\\\hline
$\frac{B(E2;2^+_2\rightarrow 0^+_2)}{B(E2;2^+_2\rightarrow 0^+_1)}$	& 2228(1046) 	& 240	& 13.5  & 7.0	& 1196 		& 0.96		\\[0.2cm]
$\frac{B(E2;4^+_1\rightarrow 2^+_2)}{B(E2;4^+_1\rightarrow 2^+_1)}$	& 1.92(14) 		& 3.0	& 0.40  & 0.012	& 185		& 0.022 		\\[0.2cm]
$\frac{B(E2;4^+_2\rightarrow 2^+_2)}{B(E2;4^+_2\rightarrow 2^+_1)}$	& 1.37(10) 		& 1.3	& 190  & 132	& 0.019		& 153		\\[0.2cm]
$\frac{B(E2;6^+_2\rightarrow 4^+_2)}{B(E2;6^+_2\rightarrow 4^+_1)}$	& 25.9(19) 		& 95	& 1477  & 1008	& 1281 		& 13704		\\[0.2cm]
$\frac{B(E2;8^+_2\rightarrow 6^+_2)}{B(E2;8^+_2\rightarrow 6^+_1)}$	& 237(47) 		& $2.1 \times 10^5$	& 4325 &   14880	& 0.29		& -		 	\\[0.2cm]
\end{tabular}
\end{ruledtabular}
\end{table}

The relative $B(E2)$ values were derived from the measured $\gamma$-ray transition intensities and compared with theory, see Tables \ref{tab:models182Hg}, \ref{tab:models184Hg} and \ref{tab:models184Hg}. The best reproduction is obtained with the IBM Phen model, however, the agreement is not as good as for the excitation energies. In particular, the $\frac{B(E2;6^+_2\rightarrow 4^+_2)}{B(E2;6^+_2\rightarrow 4^+_1)}$ ratio is overestimated in all three nuclei and the largest discrepancy, of an order of magnitude, is observed in $^{184}$Hg. In addition, the $\frac{B(E2;8^+_2\rightarrow 6^+_2)}{B(E2;8^+_2\rightarrow 6^+_1)}$ ratio in $^{186}$Hg is overestimated by three orders of magnitude. This discrepancy is related to a very small $B(E2;8^+_2\rightarrow 6^+_1)$ value predicted by the model.

The reproduction of the $B(E2)$ ratios by IBM Gogny, GBH, SCCM and BMF is in general poor. For many values, the theoretical models do not reproduce the order of magnitude of the observable. However, a comparison of the known experimental $B(E2)$ values with the theory (see Table 8 in Ref. \cite{Wrzosek-Lipska2019} and Table VII in Supplemental Material \cite{supplemental}) indicates that while the intra-band transitions are reproduced rather well, the main issue is the correct predictions of the inter-band transition strength, which can differ up to two orders of magnitude. A similar pattern can be observed in $^{188}$Hg \cite{Siciliano2020}. 

To further understand the poor reproduction of the $B(E2)$ ratios, the SCCM model Collective Wave Functions (CWF) (see Fig. \ref{fig:sccmwf}) can be analyzed. The CWF which are the weights of the intrinsic deformations in each calculated state reveal that in all three nuclei, each band has a rather constant deformation parameter. They also show that the overlap between the oblate- and prolate-deformed states is very small which can be linked to a small mixing between states exhibiting different deformation. As a result, the predicted inter-band $B(E2)$ values are too low. It should be noted that the exploratory studies of the SCCM model performed for $^{188}$Hg indicated that this behavior might be related to an absence of triaxial degrees of freedom \cite{Siciliano2020}.  

Underestimation of the inter-band transition strength by the IBM Gogny calculations was linked to the energy difference between the prolate and oblate minima on the Potential Energy Surfaces \cite{Nomura2013}. For $^{182,184}$Hg this difference is large, therefore despite the availability of the triaxial degrees of freedom, the mixing between two configurations is hindered. At the same time, for $^{186}$Hg the mixing strength was determined to be too strong for the low-lying states which might explain a systematic overestimation of the measured $B(E2)$ ratios.

\begin{table}
\caption{\label{tab:modelsrho2}Comparison of the experimental $\rho^2(E0) \times 10^3$ values in $^{182,184,186}$Hg from this work and Ref. \cite{Kibedi2022} with the theoretical models.}
\begin{ruledtabular}
\begin{tabular}{ccccccc}
$\rho^2(E0) \times 10^3$ & Exp			& IBM & IBM & GBH 	& SCCM 		& BMF	\\
& & Phen & Gogny & & & \\\hline
$^{182}$Hg($0^+_2\rightarrow 0^+_1$)	& 186 			& 34			& 8.2			& 346	& 73 		& 228	\\
$^{182}$Hg($2^+_2\rightarrow 2^+_1$)	& 110(40) 		& 196			& 1.7			& 72	& 6.7 		& 35	\\
$^{184}$Hg($0^+_2\rightarrow 0^+_1$)	& 4.1(14) 		& 31			& 4.7			& 257	& 4.5 		& 359	\\
$^{184}$Hg($2^+_2\rightarrow 2^+_1$) 	& 90(30) 		& 261			& 1.4			& 37	& 41 		& 17	\\
$^{186}$Hg($0^+_2\rightarrow 0^+_1$)	& $>50$ 		& 11			& 1.4			& 209	& 3.0 		& 120	\\
$^{186}$Hg($2^+_2\rightarrow 2^+_1$)	& 49(23) 		& 116			& 0.7			& 28	& 2.8 		& 31	\\
\end{tabular}
\end{ruledtabular}
\end{table}

The monopole strength $\rho^2(E0)$ is directly proportional to the changes in the mean-square charge radii \cite{Wood1999} and, consequently, carries important information to assess shape changes. In the case of $^{184}$Hg, we were able to reevaluate ${\rho^2(E0;0^+_2 \rightarrow 0^+_1) \times 10^3} = 4.1(14)$ by combining the intensity ratio with the known lifetime. For the $0^+_2 \rightarrow 0^+_1$ transitions in $^{182,186}$Hg and the $2^+_2 \rightarrow 2^+_1$ transitions in all three isotopes the monopole strength is known from the literature \cite{Kibedi2022}. The comparison between the experimental values and the theoretical models is presented in Table \ref{tab:modelsrho2}. 

Unlike the case of the $B(E2)$ ratios and the excitation energies, the IBM Phen predictions for the monopole strenght differ by up to one order of magnitude from the experimental data. The IBM Gogny calculations predict correctly the monopole strength between the $0^+$ states in $^{184}$Hg, however, for other analyzed cases it is underestimate by up to two orders of magnitude.

The results of GBH and BMF calculations are of the same order of magnitude. However, the monopole strength is overestimated between the $0^+$ states and underestimated between the $2^+$ states. One explanation of this effect might be an incorrect estimation of mixing between the low-spin states, as suggested in Ref. \cite{Yao2013}. The SCCM calculations are able to correctly reproduce the monopole strength in $^{184}$Hg but the predictions for $^{182,186}$Hg are too low compared to the experimental values.

It should be noted that in all discussed cases the large relative uncertainties hinder more quantitative assessment of different theoretical approaches. In addition, we bring attention to the fact that the experimental monopole strengths between the $0^+$ states in $^{182,186}$Hg might be incorrect. In case of $^{182}$Hg, ${\rho^2(E0;0^+_2 \rightarrow 0^+_1)}$ was extracted in a model-dependent way. The same approach applied to $^{184}$Hg leads to the two-orders-of-magnitude higher value than the experimental result \cite{Wrzosek-Lipska2019}. For $^{186}$Hg the method used to extract the lifetime of the $0^+_2$ state \cite{Joshi1994} suffers from unaccounted systematic effects. As shown in Ref. \cite{Olaizola2019} and discussed in details in Sec. VD therein, the same method applied to the lifetime extraction of the $2^+_2$ state in $^{188}$Hg resulted in one-order-of-magnitude difference compared to the fast-timing experiment \cite{Olaizola2019}.

\section{\label{sec:conclusions}Conclusions and outlook}

A spectroscopic study of $^{182,184,186}$Hg has been performed at the ISOLDE Decay Station at the ISOLDE facility at CERN. The excited states were populated in the $\beta$ decay of $^{182,184,186}$Tl isotopes produced in the spallation of a UC$_x$ target. The collected data allowed us to confirm the existing decay schemes and to add to them a large number of new transitions and excited states. Internal conversion coefficients were measured for 23 transitions, out of which 12 had an $E0$ component. In $^{182}$Hg, a $B(E2)$ ratio from our study combined with the results from the Coulomb excitation study allowed us to extract the sign of one interference term and to extend the systematic comparison of matrix elements with the two-state mixing model. By using electron-electron coincidences, a $0^+_3$ state was identified in $^{184}$Hg. 

The experimental results were compared with theoretical calculations. All models described qualitatively the structure of the analyzed nuclei and pointed to the coexistence of oblate- and prolate-deformed structures. However, the quantitative description is still lacking as none of the discussed approaches was able to predict correctly all the observables. A relatively good reproduction of the data was obtained in the phenomenological Interacting Boson Model with Configuration Mixing and the microscopical symmetry-conserving configuration mixing model. In particular, the latter was able to correctly reproduce the order of magnitude of the monopole strengths in $^{184}$Hg. 

The results presented in this work provide an important complementary spectroscopic input for future Coulomb excitation experiments \cite{Wrzosek-Lipska2012a,Wrzosek-Lipska2016}. They also indicate that the future experiments should focus on lifetime measurements, in particular for the low-lying yrare states, and the angular correlation to better characterize $E0$ transitions and, consequently, shape coexistence in these nuclei.

\begin{acknowledgments}

We would like to thank Tibor Kib\'edi for fruitful discussions regarding internal conversion coefficients.

We acknowledge the support of the ISOLDE Collaboration and technical teams. This project has received funding from the European Union’s Horizon 2020 research and innovation programme grant agreements No. 654002 (ENSAR2), No. 665779 and 771036 (ERC CoG MAIDEN). TRR acknowledges the computing resources and assistance provided by GSI-Darmstadt and CCC-UAM. This work has been funded by FWO-Vlaanderen (Belgium), by GOA/2015/010 (BOF KU Leuven), by the Interuniversity Attraction Poles Programme initiated by the Belgian Science Policy Office (BriX network P7/12), by the Slovak Research and Development Agency (Contract No. APVV-18-0268), by the Slovak Grant Agency VEGA (Contract No. 1/0651/21), by Spanish grants No FPA2015-64969-P, FPA2015-65035-P, FPA2017-87568-P, FPA2017-83946-C2-1-P, RTI2018-098868-B-I00, PPID2019-104002GB-C21, PID2019-104390GB-I00, PID2019-104714GB-C21, and PID2019-104002GB-C21 funded by MCIN/AEI/10.13039/50110001103 and “ERDF A way of making Europe” and by ERDF, ref. SOMM17/6105/UGR, by Science and Technology Facilities Council (STFC) of the UK Grant No. ST/R004056/1, ST/P005314/1, ST/P003885/1, ST/V001035/1, ST/P004598/1 and ST/V001027/1, by the German BMBF under contract 05P21PKCI1, by the Romanian IFA project CERN-RO/ISOLDE, by the Polish Ministry of Education and Science under the contract nr. 2021/WK/07 and by the Academy of Finland (Finland) Grant No. 307685.

The experimental data used in this study are available from the corresponding author upon reasonable request.

\end{acknowledgments}

\bibliographystyle{apsrev}
\bibliography{Tlbetaref,other}

\end{document}